\documentclass[11pt]{iopart}
\usepackage{iopams} 
\usepackage{graphicx,epsf,color}

\newif\ifpreprint
\preprinttrue

\def\f{\tilde f}

\def\NeqFoursYM{{${\cal N}=4$~sYM}}
\def\NeqFour{{{\cal N}=4}}

\def\NeqEight{{{\cal N}=8}}

\def\Neqone{{{\cal N}=1}}
\def\tree{{\rm tree}}
\def\be{\begin{equation}}
\def\ee{\end{equation}}
\def\bea{\begin{eqnarray}}
\def\eea{\end{eqnarray}}
\newcommand{\nn}{\nonumber}
\def\tree{{\rm tree}}

\def\tag#1{\tree}

\def\spa#1.#2{\left\langle#1\,#2\right\rangle}
\def\spb#1.#2{\left[#1\,#2\right]}
\def\tlambda{\widetilde{\lambda}}
\def\PlusVertex{$(+)$}
\def\MinusVertex{$(-)$}

\def\fig#1{figure~{\ref{#1}}}
\def\Fig#1{Figure~{\ref{#1}}}
\def\eqn#1{(\ref{#1})}
\def\Eqn#1{Equation~(\ref{#1})}
\def\sect#1{section~{\ref{#1}}}
\def\sv{s_{12}}
\def\tv{s_{23}}

\def\UniversalFactor{U}

\begin{document}
\ifpreprint
SU-ITP-11/07 $\null\hskip0.2cm\null$ \hfill 
\hbox{~~~~}Saclay--IPhT--T11/029
\fi

\title{Generic multiloop methods and \\ application to 
${\cal N}=4$ super-Yang-Mills}

\author{John~Joseph~M.~Carrasco${}^a$ and Henrik~Johansson${}^b$ }

\address{
${}^a$Stanford Institute for Theoretical Physics and Department of Physics, Stanford University,
             Stanford, CA 94305-4060, USA \\
${}^b$Institut de Physique Th\'eorique, CEA--Saclay,
          F--91191 Gif-sur-Yvette cedex, France
}
	
\ead{jjmc@stanford.edu, henrik.johansson@cea.fr}
\date{\today}

\begin{abstract}
We review some recent additions to the tool-chest of techniques for finding compact integrand representations of multiloop gauge-theory amplitudes --- including non-planar contributions --- applicable for $\NeqFour$ super-Yang-Mills in four and higher dimensions, as well as for theories with less supersymmetry.   We discuss a general organization of amplitudes in terms of purely cubic graphs, review the method of maximal cuts, as well as some special $D$-dimensional recursive cuts, and conclude by describing the efficient organization of amplitudes resulting from the conjectured duality between color and kinematic structures on constituent graphs.  

This article is an invited review for a special issue of Journal of Physics A devoted to ``Scattering Amplitudes in Gauge Theories''.
\end{abstract}

\clearpage
\tableofcontents
\clearpage
\markboth{Generic multiloop methods and application to 
${\cal N}=4$ super-Yang-Mills}{}

\section{Introduction}
Over the past few years there has been spectacular progress~\cite{MHVoneloop, NMHVoneloop, BCDKS, FiveLoop, KorchemskyOneLoop,TwoLoopSixPt, CachazoLeadingSingularityAndCalcs, Spradlin3loop,  VerguAnd2L6pNMHV, Grassmannians, AHIntegrands,TwistorWilsonLoopAndDualCSW} in writing down the integrands of loop-level scattering amplitudes of the ${\cal N}=4$ super Yang-Mills (sYM) theory~\cite{NeqFourDefinition} in the planar leading-color sector, or 't Hooft limit. Similarly, for the integrated planar multiloop amplitudes there are recent important advances, notably the all-loop resummation of four and five-particle amplitudes proposed by Anastasiou, Bern, Dixon, Kosower, and Smirnov~\cite{BDS}, and by the fully analytic evaluation of the six-particle two-loop maximally-helicity-violating amplitude~\cite{AnalyticSixPoint}; beyond that, there are many hints, both at weak and strong coupling~\cite{MagicIdentities, AldayMaldacena, AmplitudeWilsonLoopDuality,  DualSuperconformalSymmetry, Yangian, MassiveRegulatorProgress, LoopPolytopes, BeisertProgram, correlatorsAndamplitudes}, that the planar sector may soon be solved.
  
The subleading color contributions to \NeqFoursYM\ amplitudes are less well studied---save for the one loop case, where all subleading-color amplitudes may be obtained from the planar sector ones (see e.g. \cite{MHVoneloop,treeLoopcolorDecomp}).  Beyond one loop, the literature contains only the full non-planar contributions of the four-point amplitude, worked out at two~\cite{BRY}, three~\cite{superfinite, CompactThree,BCJloops} and four~\cite{Neq44np} loops at the level of the integrand.  Notably, these four-point amplitudes also happen to be some of the few amplitudes known to be valid in higher dimensions $D>4$ \cite{Neq44np,SixDim} in the dimensionally oxidized ${\cal N}=4$ theory (dimensional reductions of the $D=10$ sYM theory). 

In this contribution to the review, we will discuss some generally applicable methods for constructing multiloop multiparticle integrands of amplitudes, valid for \NeqFoursYM\ in four and higher dimensions, including non-planar contributions. Much of what is covered also applies to less supersymmetric theories, including pure QCD; however, the focus and examples in this chapter will be on the \NeqFoursYM\  theory.  We will exhibit the general organization of unitarity cuts and amplitudes, specifically the maximal cut method~\cite{FiveLoop,CompactThree,Neq44np} and the imposition of a duality between color and kinematic structures of amplitudes~\cite{BCJ,BCJloops}.  Interestingly, the duality allows one to recognize that all contributing  terms of the amplitude are fully specified by some small set of independent ``master graphs''.  A discussion which only applies to \NeqFoursYM\ is the review of special recursive cuts  known as ``two-particle cuts''~\cite{BRY,BDDPR,Neq44np}, and ``box cuts''~\cite{Neq44np}, relying on the fact that all four-point amplitudes in \NeqFoursYM~factorize into a product of the corresponding tree amplitude times a universal function.

Loop amplitudes of massless theories in four dimensions suffer from infrared divergences (and frequently also ultraviolet divergences). A standard approach for avoiding these infrared divergences is to dimensionally regulate the amplitudes, that is, to work in $D=4-2\epsilon$ with $\epsilon<0$.
In this scheme, it is thus critical to compute loop amplitudes in $D>4$ dimensions. The methods presented in this review are well suited for this task.
 
Throughout this contribution we make use of the unitarity method~\cite{UnitarityMethod} and generalized unitarity~\cite{GeneralizedUnitarity,GeneralizedUnitarity2,BCF}, which we consequently will briefly outline.  However, deeper understanding of details, as well as applications of this approach, may be found in some of the parallel chapters of this review, {\it e.g.} Bern and Huang~\cite{ZviYutinReview}, Britto~\cite{BrittoReview}, and Ita~\cite{ItaReview}.
 
The organization is as follows. In \sect{PreliminariesSection}, we set up notation and organization of multiloop integrands and cuts.   In \sect{MaximalCutsSection}, the method of maximal cuts is discussed in some detail, focusing on both the systematic identification of needed cuts as well as the computation of these using tree amplitudes as input. Complimentary to the method of maximal cuts, in \sect{RecursiveCutsSection}, we review two particularly simple classes of multiloop cuts, relevant to the \NeqFoursYM\ theory:  {\it two-particle cuts}, and {\it box cuts}.  Finally, in \sect{BCJ}, we discuss the set of powerful constraints that one can impose on multiloop gauge theory amplitudes by considering the color--kinematics duality.

\section{Preliminaries}
\label{PreliminariesSection}

\subsection{Integral representations for amplitudes}

To systematically employ the unitarity method~\cite{UnitarityMethod, GeneralizedUnitarity,GeneralizedUnitarity2}, it is convenient to organize the amplitude using a sufficiently general integral representation.
We note that one can express any local massless gauge theory amplitude in $D$ dimensions in terms of local interactions and non-local propagators $\sim 1/p^2$, regardless of the particle content ({\it cf.} Feynman gauge propagators and vertices).   With this in mind, one can write any amplitude as
\be
{\cal A}_n^{(L)} \sim \sum_{i\in\Gamma} \, \int \frac{d^{LD}\ell}{(2\pi)^{LD}}  \, \frac{\rm local}{p_{i_1}^2p_{i_2}^2p_{i_3}^2\cdots p_{i_m}^2}\,,
\label{localRep}
\ee
where $d^{LD}\ell=\prod_{j=1}^{L}d^D\ell_j$ is the usual Feynman diagram integral measure of $L$ independent $D$-dimensional loop momenta $\ell_j^\mu$, and $\Gamma$ is the set of all $L$-loop graphs, with $n$ external legs, counting all relabeling of external legs. Corresponding to each internal line (edge) of the $i^{\rm th}$ graph, we associate a propagator $1/p_{i_l}^2$, which is a function of the independent internal and external momenta, $\ell_j$ and $k_j$, respectively.  The local numerator functions, here only schematically indicated, include information about the color, kinematics and states, {\it etc}.  This representation thus mimics the behavior of Feynman diagrams, but as we will see the local numerators need not be calculated using off-shell Feynman rules.  Instead it is often more efficient to indirectly infer these numerator functions by imposing general constraints and by matching to physical on-shell information gleaned from generalized unitarity cuts. 

Clearly the way of expressing an amplitude in \eref{localRep} is not unique.  One can always add more propagators to any graph by multiplying by unity, expressed as $p_{i_l}^2/p_{i_l}^2$, absorbing the numerator $p_{i_l}^2$ into the graph's local numerator factor without changing its locality. To remove this freedom, we can insist that the graphs we use have only cubic (trivalent) interactions. That is, we can hide the contact terms that one usually encounters in gauge theories through the repeated multiplication of $p_{i_l}^2/p_{i_l}^2$ factors until every graph has the same number of propagators as a cubic graph: $n+3L-3$. Thus, we use a more constrained representation of the amplitude,
\be
{\cal A}_n^{(L)}=\sum_{i\in\Gamma_3} \, \int \frac{d^{LD}\ell}{(2\pi)^{LD}} \frac{1}{S_i} \, \frac{N_i C_i}{p_{i_1}^2p_{i_2}^2p_{i_3}^2\cdots p_{i_{m}}^2}\,,
\label{cubicRep}
\ee
where $\Gamma_3$ is the set of all purely cubic graphs, and $m=n+3L-3$ for all graphs. Furthermore, we have, for convenience, separated out the kinematic parts $N_i$ and the color parts $C_i$ of the local numerator factor, and normalized the numerators by dividing out by the standard (Bose) symmetry factor $S_i$ of each graph.  

One might well wonder if it is always possible to factorize the kinematic and color terms for each graph.  Indeed, this requirement imposes severe constraints on the local numerator.  Nonetheless, when all particles are in the adjoint representation of the gauge group, as is the situation for the theories we will consider,  this factorization is possible~\cite{BCJ}. It can be seen from the Feynman rules of gauge theory: if all particles are in the adjoint the only color structure that can appear in cubic vertices are the gauge group structure constants $f^{abc}$. Hence, every purely cubic Feynman diagram has a unique color factor associated to it. For the quartic contact vertices of gauge theory, there are three different color structures: $f^{a_1a_2b}f^{ba_3a_4}$ and its permutations. If we promote such a contact term to be a part of a cubic graph using $f^{a_1a_2b}f^{ba_3a_4} \rightarrow p^2_{12}/p^2_{12}f^{a_1a_2b}f^{ba_3a_4}$, then this term is proportional to the same color factor as the non-contact contribution in the $p^2_{12}=(p_1+p_2)^2$ channel. Thus, by this construction, the color factor of each cubic graph in \eref{cubicRep} is unique, up to a normalization factor. They are given by
\be
C_i=\prod_{j\in V_{i}}\f^{a_{j_1}a_{j_2}a_{j_3}}\,,
\label{ColorDef}
\ee
where the product runs over all the cubic vertices in the $i^{\rm th}$ graph. Here, we use structure constants $\f^{abc}$ with a normalization differing from normal conventions, $\f^{abc}= i \sqrt{2} f^{abc} = \Tr([T^a, T^b] T^c)\,$, and with hermitian generators $T^a$ normalized via $\Tr(T^a T^b) = \delta^{ab}$. 

The kinematic local numerators $N_i$ are by definition polynomials in the loop momenta but may be taken to be either polynomials or rational functions of external kinematic variables depending on the convenience of the situation at hand. Needless to say, $N_i$ also depends on the polarizations, helicities and spinors, {\it etc}, according to the details of the external states. The maximal degree of the polynomial in loop momenta can be determined by dimensional analysis: an $n$-particle gauge theory amplitude in $D=4$ has dimension $4-n$; thus, after accounting for the dimension of the integral measure and propagators, $N_i$ has dimension $n+2L-2$. This dimensionality coincides with the number of cubic vertices in each graph, as is expected of a gauge theory where each cubic vertex has one power of momenta. The dimension of $N_i$ then has to be distributed between the factors of external and internal momenta. This means that in a generic gauge theory, one would expect that the loop momentum dependence in $N_i$ comes in monomials of the form
\be
1, \ell^{1}, \ell^{2},\ldots, \ell^{n+2L-3},\ell^{n+2L-2}~~ \in~~ N_i\,,
\label{monomials}
\ee
where $\ell^{p}$ is a mnemonic for any Lorentz tensor constructed out of $p$ factors of loop momenta $\ell_j^\mu$. For special gauge theories, the highest degree monomials are absent for all $N_i$. This property can be attributed to the good ultraviolet behavior of the theory, as the ultraviolet regime is indeed the corner of phase space in which $\ell_j^\mu \rightarrow \infty$. 

Although the ultraviolet behavior of a theory is interesting in itself, for the purposes of this review we will only note that such behavior  constrains the form of the amplitude.  In the absence of other constraints, the fewer powers of $\ell$ that one must consider, the smaller the space of possible numerator polynomials will be.  This can greatly facilitate amplitude construction. The set of well-known gauge theories with good ultraviolet behavior includes many supersymmetric Yang-Mills theories. In particular, maximally supersymmetric  \NeqFoursYM\ is known to reduce the maximum power of the monomials \eref{monomials} by at least four relative to pure Yang-Mills theory~\cite{NeqFourFiniteness,MS,BDDPR,HoweStelleRevisited}.  Theories with ${\cal N}$-fold supersymmetry are  expected to reduce the maximal power of $\ell$ by at least ${\cal N}$ in \eref{monomials}, see {\it e.g.}~\cite{UnexpCanc}. In some cases, the cancellations of the high powers of $\ell$ are even stronger than this naive counting~\cite{MS,Neq44np, DoubleTraceArguments}.

For completeness, we also mention an alternative integral representation frequently used in the literature. For amplitudes with many external legs (compared to the number of spacetime dimensions), it can be cumbersome to use an integral representation with only cubic vertices. For example, the standard one-loop integral basis in four dimensions consists of box, triangle and bubble integrals; no higher polygon integrals are needed for massless $D=4$ theories (see {\it e.g.}~\cite{BrittoReview}). A similar basis is expected to exist at the two-loop level~\cite{Kosower2Lbasis}. To allow for such cases, we use a sufficiently general (over-complete) integral representation
\be
{\cal A}_n^{(L)}=\sum_{k\in {\rm TB}}G_{k} \sum_{i\in\Gamma} \, R_{i,k} \int \frac{d^{LD}\ell}{(2\pi)^{LD}} \,  \frac{P_{i,k} }{p_{i_1}^2p_{i_2}^2p_{i_3}^2\cdots p_{i_{m}}^2}\,,
\label{convRep}
\ee
where the first sum runs over color factors $G_k$ in the trace basis (TB), of say $SU(N_c)$, and the second sum runs over $n$-point $L$-loop graphs as in \eqn{localRep}. The $R_{i,k}$ are rational functions of external momenta, and $P_{i,k}$ are polynomials of the independent loop momenta. As desired, one can generally restrict the number of propagators $m \le  \lceil D\rceil  L$ ($\lceil \bullet \rceil$ is the ceiling function),  because integrals with $m>\lceil D\rceil L $ are reducible ({\it e.g.} using Melrose, Van-Neerven-Vermasseren, or Bern-Dixon-Kosower reduction and similar methods~\cite{integralReductionsMNVBDK}), although the exact upper bound depends critically on the subtleties of dimensional regularization at each loop order. Similar to the representation \eref{cubicRep} the ultraviolet behavior of the theory under consideration imposes useful constraints on the form of $P_{i,k}$. For example, the good ultraviolet behavior of \NeqFoursYM\ can be used to show that only scalar boxes contribute at one loop in $D=4$ \cite{MHVoneloop}; thus, one can set $P_{i,k}=1$ for all boxes and $P_{i,k}=0$ for all other one-loop integrals in this theory. 

As a recent alternative to the mentioned general integral representations, we remark on an interesting new set of chiral integrals that have been introduced specifically for planar \NeqFoursYM\ using momentum twistor space notation~\cite{AHIntegrands}.  Indeed, using special properties of a theory, such as dual (super-)conformal symmetry~\cite{MagicIdentities, DualSuperconformalSymmetry} (or Yangian structure~\cite{Yangian}) of \NeqFoursYM, one may construct more constrained integral representations, which consequently increases the power of any amplitude calculation~\cite{BCDKS, FiveLoop, KorchemskyOneLoop,TwoLoopSixPt}. In similar spirit, the duality between color and kinematics~\cite{BCJ,BCJloops} discussed in \sect{BCJ}, is conjectured to provide an constraining organizing principle for the integrals in more general gauge theories. 

We note that the color--kinematics duality will critically rely on the cubic integral representation \eref{cubicRep}, whereas in sections \ref{MaximalCutsSection} and \ref{RecursiveCutsSection}, which treat generalized unitarity cuts, either integral representation  \eref{cubicRep} or \eref{convRep} is applicable.

\subsection{Generalized unitarity cuts}

Unitarity of the $S$-matrix requires that loop amplitudes must satisfy certain properties relating them to sums over products of lower-loop amplitudes.  The goal of the unitarity method~\cite{UnitarityMethod, GeneralizedUnitarity,GeneralizedUnitarity2}  is to construct a compact expression that obeys all constraints required by unitarity for a particular amplitude. For massless theories it is sufficient to satisfy all generalized unitarity cuts in $D$ dimensions, this guarantees that no logarithmic or rational terms can go undetected. Strictly speaking only a subset of these, a "spanning set" of cuts~\cite{Neq44np}, are necessary to guarantee the satisfaction of all other cuts.

To identify contributions to the amplitude, we will compute generalized unitarity cuts on the level of the integrand, expressing them as a product of on-shell subamplitudes,
\be
i^c \sum_{\rm states}A_{(1)}\,A_{(2)}\,A_{(3)}\cdots A_{(m)}\,,
\label{cutdef}
\ee
where $A_{(i)}$ are the subamplitudes, which may be either tree or lower-loop amplitudes. The number of cut lines for each cut (the number of on-shell $p_i^2=0$ internal legs) is denoted by $c$. 

The cuts can be computed directly from the theory by feeding in the corresponding subamplitudes and summing over the intermediate states, as in \eref{cutdef}. They may, of course, also be computed from graph-based integrand representations of the full loop amplitude, \eref{cubicRep} or \eref{convRep}, when available.  In this latter case, the sum over intermediate states has already been carried out, and the generalized cut can be formally obtained by replacing the propagators of the cut lines by delta functions,
\be
\frac{1}{p_i^2} \rightarrow 2\pi \delta(p_i^2)\,,
\label{cutdef2}
\ee
and collecting the terms with the most delta functions. Since the full integrand of the amplitude under consideration is typically {\it a priori} unknown, the main purpose of these types of cuts is to apply them to an ansatz of the amplitude.

The unitarity method instructs us to reconstruct amplitudes using the information from the set of all generalized unitarity cuts of a theory. This set is overcomplete; consequently, there are different strategic approaches on how to most efficiently extract the relevant information for building the amplitudes. For example, one might well consider a strategy of using the fewest number of cuts, with each individual cut thereby providing the most information about a given amplitude.   The most extreme such situation is exemplified by the one-particle cut of an $n$-point $L$-loop amplitude, relating it to a $(n+2)$-point $(L-1)$-loop amplitude sewn with itself.  This cut has the minimal number of cut propagators, and in principle contains the full information of all the other generalized cuts of the $L$-loop amplitude (ignoring any subtleties with possible collinear and soft divergences of the cut line).   Without an organizing principle, disentangling this information can be prohibitive.   In the planar-limit of~\NeqFoursYM, dual conformal symmetry (or Yangian invariance) appears to provide such an organizing principle~\cite{AHIntegrands}.  See also~\cite{BrittoSingleCut} for a general discussion of single-line cuts at the one-loop level.

In the next section, we consider a different strategy, using the method of maximal cuts.  Rather than maximizing the information coming from each cut, we maximize the number of cuts and similarly the number of cut lines, ensuring that each cut provides as small an amount of information as possible, thus facilitating the piece-by-piece construction of the amplitude.

%%%%%%%%%%
\section{Maximal cuts in $D=4$ and general dimension}
\label{MaximalCutsSection}

The method of maximal cuts~\cite{FiveLoop,CompactThree,Neq44np} offers a particularly efficient means for determining multiloop amplitudes.   In this method, we start from those generalized cuts that have the maximum number of cut propagators (maximal cuts). These cuts provide an initial ansatz for the amplitude. Next we systematically correct the ansatz by considering the information provided by near-maximal cuts. These sequentially reduce the number of cut propagators one by one. In this process, all potential contact contributions are determined.   

Generalized unitarity has a wide range of applicability: amplitudes of generic theories in generic spacetime dimension can be constructed.  Since maximal and near-maximal cuts are special cases of generalized unitarity cuts, they can be used in many theories and space-time dimensions. However, there are some mild assumptions necessary if one wishes to organize around cubic vertices and cut all propagators:  it is important that massless on-shell three-point amplitudes are non-vanishing and non-singular~\cite{BCF,BuchbinderCachazo}, for appropriate choices of complex cut loop momenta~\cite{GoroffSagnotti,WittenTopologicalString}. Moreover, it is important that the space-time is of dimension $D\ge 4$, in order to allow for one to impose both momentum conservation and on-shell conditions of the legs meeting at a three-point vertex without having collinear momenta\footnote{These assumptions are needed if one wants to cut down to cubic vertices.  In theories that have no on-shell three point vertices amplitudes may instead be organized around quartic vertices, which has no such restrictions~\cite{ThreeDimAmpl}.}.

The strategy of maximizing the number of cut propagators is known to be a powerful method for computing one-loop \NeqFoursYM\ amplitudes in four dimensions, because
only box integrals appear~\cite{MHVoneloop}.  As observed by
Britto, Cachazo and Feng~\cite{BCF}, taking a quadruple cut,
where all four propagators in a box integral are cut, freezes the
four-dimensional loop integration.  This allows the kinematic coefficient of the box
to be straightforwardly computed directly in terms of the cut.  The use of complex
momenta, as suggested by twistor space
theories~\cite{WittenTopologicalString}, makes it possible to define
massless three-vertices and thereby to use quadruple cuts to determine
the coefficients of all box integrals including those with massless
external legs.

The idea behind the quadruple cuts has been generalized by Buchbinder and Cachazo~\cite{BuchbinderCachazo} to two loops using hepta- and
octa-cuts of the double box integral topology. Although the two-loop four-point double-box integral only has seven physical propagators, it secretly enforces an additional spurious eighth constraint, yielding an octa-cut which localizes the integration
completely.  These types of cuts are now known as a part of the ``leading-singularity''
technique, which is similar to the maximal cut method in that it is based on cutting 
a maximal or near-maximal number of
propagators~\cite{CachazoSkinner,CachazoLeadingSingularityAndCalcs},
but in addition it may make use of extra conditions from spurious
singularities that are special to four dimensions.

The strength of the method of maximal cuts has most clearly been seen in amplitudes that have a large number of multiloop integrals. For example, the four-point amplitudes of \NeqFoursYM\ have been evaluated using maximal and near-maximal cuts at five loops~\cite{FiveLoop} in the planar sector, and similarly for the full non-planar amplitudes at three~\cite{CompactThree} and four loops~\cite{Neq44np}, the latter containing 50 distinct integrals, each with multiple terms.

\subsection{Working definition of maximal and near-maximal cuts}
\label{maxcutdefSection}
To make the concept of maximal cuts clear, we will introduce some proper definitions.  First we note that one can represent generalized cuts~\eref{cutdef} as graphs: vertices (nodes) representing on-shell subamplitudes, and lines (edges) representing on-shell propagators. Maximal cuts are simply those  generalized cuts (graphs) that have the most possible on-shell cut conditions, given three numbers: $n$  the number of external legs, $L$ the number of loops and $D$ the number of dimensions of the loop momenta. For any generalized cut, there are two competing upper limits to the number of cut conditions $c$ that one can impose:
\bea
c\le n+3L-3\,, \hskip 1cm {(\rm cubic~graph~limit)} \nn \\
c\le DL\,, \hskip 1cm~~~~~~~~~~(\mbox{loop~momentum~freeze-out})
\label{MCbound}
\eea
where the first limit comes from counting the internal lines of a cubic $n$-point $L$-loop graph, and the second by counting the degrees of freedom of the independent loop momenta. It is clear that a cut consisting of only cubic vertices can have no more physical singularities as the integrand is completely local. Similarly, a cut where all the degrees of freedom of the loop momenta are frozen can have no more singularities in the internal phase space.

We define maximal cuts to be those generalized cuts which saturate one of the above bounds\footnote{In some unitarity cuts, the ``loop momentum freeze-out'' bound, which counts the total degrees of freedom of the loop momenta, is not a good measure of what happens locally in each subgraph. For example, a direct product of a one-loop triple cut and a one-loop penta-cut would be a two-loop maximal cut in four dimensions, since it is an octa-cut. However, since the one-loop penta-cut does not exist in four dimensions, one could argue that it should not be counted. Nonetheless, in order to simplify our definition~\eref{MCdef}, we simply count such situations as valid cuts that happens to evaluate to zero.}, that is,
\be
c_{\rm MC}={\rm min}\{n+3L-3,DL\}\,.
\label{MCdef}
\ee
For example, for $D=4$ and $L=1$ the maximal cut is equal to the quadruple cut $c_{\rm MC}=4$. For $n=4$ and $L=2$ it is the hepta-cut $c_{\rm MC}=7$, {\it etc}. (For comparison, we note that if we work in four dimensions, then for loop orders $L \le n-3$ the maximal cuts coincide with the leading singularity cuts as both have $4L$ cut lines.)

To define the set of near-maximal cuts, we consider those cuts with fewer cut lines than the maximal cuts $c<c_{\rm MC}$. In particular, a next-to-maximal cut has $c=c_{\rm MC}-1$ cut lines, and a next-to-next-to-maximal cut have  $c=c_{\rm MC}-2$ cuts, and so on. For example, in this classification, we refer to both one-loop triple cuts in $D=4$ and quadruple cuts in $D=5$ as next-to-maximal cuts (for $n>4$).

\subsection{An algorithm for the maximal cut method}

We give here a general outline for constructing the loop amplitudes using the method of maximal cuts. 
In words, the algorithm is as follows:

\begin{enumerate}
\item Enumerate all possible maximal cuts given the multiplicity $n$, loop order $L$, and dimension $D$. For example, if $n\le D$, or more generally $L>(n-3)/(D-3)$, the maximal cuts are in one-to-one correspondence with the cubic graphs of loop order $L$ and multiplicity $n$ (as follows from \eref{MCdef}).
\item Compute the maximal cuts at the level of the integrand ({\it i.e.} \eref{cutdef}). If the maximal cuts do not freeze the loop momenta, make sure to capture this residual dependence using fully Lorentz covariant variables of the loop momenta (Lorentz products or Levi--Civita tensor contractions).
\item Using the information from the maximal cuts, construct an initial ansatz for the amplitude. By the Lorentz covariant construction, the expressions obtained from the maximal cuts are automatically pushed to off-shell loop momenta (the validity of any such continuation will be ensured by the consecutive near-maximal cuts).   
\item Enumerate  the next-to-maximal cuts.  Calculate each cut using \eref{cutdef}, then compare to the result of each cut as obtained from the amplitude ansatz constructed in the previous step. If the difference between the true cut and the cut of the ansatz is nonzero, then correct the ansatz by incorporating said difference.  
\item Continue iterating the previous step for the next-to-next-to-maximal cuts, and so on; each step with one fewer cut condition imposed. 
\end{enumerate}

The iteration can be stopped at any point when the amplitude ansatz appears to receive no more corrections. A proof that the ansatz is  correct will follow from evaluating it on a complete set of ``spanning cuts''~\cite{Neq44np}, which are cuts that are known to fully determine $n$-point $L$-loop amplitudes of any theory (see~\cite{ZviYutinReview} for more details). Alternatively, the iteration may be continued by removing cut conditions one by one until the iteration naturally terminates by generating its own set of complete ``spanning cuts''. 

If we represent the amplitude using the integrals in \eqn{cubicRep}, then the amplitude is entirely specified by the integral numerators $N_i$. The idea of the above algorithm is then that the numerators can be expanded as\footnote{It should be noted that this expansion is not unique. For example, contributions of the amplitude corresponding to contact terms can be shuffled between different graphs without affecting the amplitude, nor the generalized unitarity cuts. The algorithm of the maximal cut method simply gives a representation of the amplitude that satisfies all unitarity cuts; thus, much of the reparametrization freedom of the full amplitude is present in this representation.}
\be
N_i=N_i^{\rm MC}+N_i^{\rm NMC}+N_i^{\rm N^2MC}+\ldots\,,
\label{expansion}
\ee
where $N_i^{\rm MC}$ is the initial ansatz determined by the maximal cuts, and the corrections from the near-maximal cuts take the schematic form
\be
N_i^{\rm NMC}=\sum_{j=1}^{m} p^2_{i_j} P_{i;j}\,, \hskip 5mm N_i^{\rm N^2MC}=\sum_{j,k=1}^{m} p^2_{i_j}p^2_{i_k} P_{i;j,k}\,, \hskip 5mm \cdots\,,
\ee
where the $p^2_{i_j}$ are inverse propagators and $P_{i;j}$ and $P_{i;j,k}$ are polynomials fixed by the next-to- and next-to-next-to-maximal cuts, respectively. The contributions from next-to-maximal cuts must be multiplied by at least one inverse propagator since these contributions must vanish on maximal cuts where all $p^2_{i_j}=0$. Similarly, the corrections from the next-to-next-to-maximal cuts be multiplied by at least two inverse propagators since they vanish on all next-to-maximal cuts, and so on. Furthermore, since the inverse propagators carry dimension, the dimension (or degree) of the local polynomials must reduce for each iteration step, yet by locality remain non-negative. This is why the algorithm naturally terminates.

In the following sections, we will review how to compute maximal cuts using gauge theory tree amplitudes as input.

\subsection{Cuts with on-shell three-vertices}

In this part, we direct our attention to the evaluation of the maximal cuts. First, we will consider maximal cuts consisting of only three-point tree amplitudes
\be
i^c \sum_{\rm states}A^{\tree}_{3,(1)}\,A^{\tree}_{3,(2)}\,A^{\tree}_{3,(3)}\cdots A^{\tree}_{3,(m)}\,,
\ee
where the number of cut conditions is given by $c=n+2L-2$. Many of the interesting features of this cut are exposed by studying the kinematics of a single on-shell tree $A^{\tree}_{3}$.

\begin{figure}
\centerline{\epsfxsize 3.5 truein \epsfbox{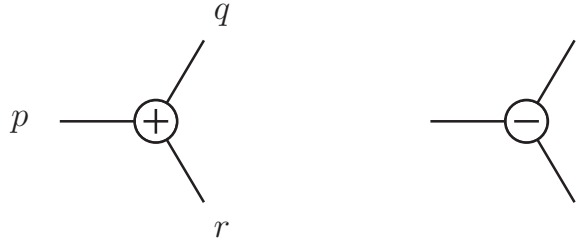}} \caption{In four dimensions, the kinematics of the on-shell three-point vertex comes in two chiral classes: the \PlusVertex\ vertex is supported on $\lambda$s (thus all $\tlambda$s are parallel and their $\spb{i}.{j}$ vanish), and the \MinusVertex\ vertex is supported on $\tlambda$s (thus all $\lambda$s are parallel and their $\spa{i}.{j}$ vanish). } \label{ChiralVertexFigure}
\end{figure}

Consider three massless momenta $p$, $q$ and $r=-p-q$ meeting at a three-point vertex, as in \fig{ChiralVertexFigure}; one has
the following constraints:
\be
 p^2=q^2 =p\cdot q=0\,,
\label{pqcond}
\ee
where the orthogonality $p\cdot q=0$ follows from the masslessness of the third leg $r^2=(p+q)^2=2p\cdot q=0$.  For real momenta in Minkowski signature $(1,D-1)$, these equations have only degenerate solutions where the momenta are collinear, $p\,\|\, q$.  For nondegenerate solutions we may either work in $(2,D-2)$ signature, or more generally, we may explicitly complexify the momentum space (the $(2,D-2)$ signature can be thought of as a special case of complexfied momenta). For example, an explicit nondegenerate solution can, choosing a convenient Lorentz frame, take the form\footnote{Given the two $D$-dimensional momenta $p$ and $q$ that satisfies \eref{pqcond}, we introduce a massive reference vector $v$ such that $v\cdot q= 0$ and $v\cdot p\neq 0$, which can always be found when $p$ and $q$ are not collinear.  By a Lorentz transformation, we may bring $v$ and $p$ to the forms $v^\mu=(\bullet,0,\ldots,0)$ and $p^\mu=(\bullet,\bullet,0,\ldots,0)$. From the conditions $v\cdot q= 0$ and $p\cdot q= 0$, we conclude that in this frame $q^\mu=(0,0,\bullet,\bullet,\ldots,\bullet)$. Next we perform a Lorentz transformation in all but the first two spacetime directions, which leaves $p$ unaffected and brings $q$ to the form $q^\mu=(0,0,\bullet,\bullet,0, \ldots,0)$.}
\be
p^\mu=(1,1,0,0,0,\ldots,0)\,, \hskip 1cm q^\mu=(0,0,1,i,0,\ldots,0) \,.
\label{embededPQ}
\ee
Interestingly the solutions in $D<4$ are strictly degenerate; thus the on-shell three-vertex is only supported by kinematics in $D\ge 4$. The case $D=4$ is the most interesting, not only because it is the spacetime we notice in our day-to-day lives but also because it is the marginal case where the phase space of the kinematics is the most constrained, yet nondegenerate. We will see the consequences of this below.

In four dimensions, we use the standard spinor helicity notation (see {\it e.g.}~\cite{DixonTASI,OnShellMethodsReview}) to solve kinematic constraints of the three-point vertex~\cite{BCF}; the momentum bispinor form $p=\lambda_p\tlambda_p$ and $q=\lambda_q\tlambda_q$ ensures the null conditions, and the orthogonality condition becomes
\be
0 = 2p \cdot q = \spa{p}.{q} \spb{q}.{p} \,.
\label{oscond}
\ee
For real momenta in Minkowski signature, $\lambda_{p}$ and ${\tlambda}_{p}$ are complex conjugates of
each other (up to a sign).  Hence, if $\spa{p}.q$ vanishes then
$\spb{p}.q$ must also vanish and the momenta are collinear.  If the momenta are taken to be complex, however, the two
spinors $\lambda_{p}$ and ${\tlambda}_{p}$ are independent. This
gives two independent solutions to \eqn{oscond},
\be
 \spa{p}.q=0 \hskip 1 cm   \hbox {\rm or} \hskip 1 cm 
\spb{p}.q=0 \,,
\label{TwoBranches}
\ee
implying that $\lambda_{p}$ and $\lambda_{q}$ are proportional or, in the latter case, that $\tlambda_{p}$ and $\tlambda_{q}$ are proportional. Since $r=\lambda_r\tlambda_r=-\lambda_p\tlambda_p-\lambda_q\tlambda_q$, it then follows that there are overall two possible solutions:
all $\lambda$s proportional, and hence all $\spa{i}.{j}$ vanishing, or all $\tlambda$s proportional, and hence all $\spb{i}.j$ vanishing. We conclude that in $D=4$, the momentum-space support of the three-vertex exhibits a remarkable twofold branching or chirality. We will denote the vertices that are supported by $\lambda$ kinematics by $(+)$ and the vertices supported by $\tlambda$ kinematics by $(-)$, as illustrated in \fig{ChiralVertexFigure}.

%%%%%%%%%%%%%%%%%%%%
%FIGURE
%%%%%%%%%%%%%%%%%%%%%%

%%%%%%%%%%%%%%%%%%%%
%FIGURE
\begin{figure}
\centerline{\epsfxsize 2.7 truein \epsfbox{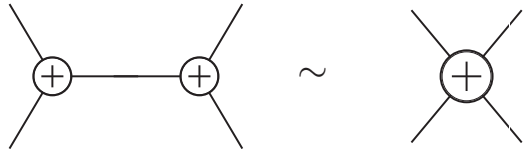}} \caption{In a $D=4$ cut, one can encounter two three-vertices of the same chirality joined by a common line ({\it e.g.} see cuts in \fig{PentaBoxFigure}(b) and \ref{non-planarFigure}). For the purpose of solving the kinematics, one can treat this configuration as an effective four-point vertex of definite chirality (note that a corresponding amplitude is not defined). All intermediate channels of this effective kinematic vertex are on-shell, as all legs are proportional to the same spinor.} \label{ChiralProductFigure}
\end{figure}
%%%%%%%%%%%%%%%%%%%%%%

The chiral branching implies that the on-shell three-point amplitudes of {\it any massless theory} in $D=4$ have the following kinematic dependence:
\bea
A_3^{(+)} \equiv A_3(\lambda_1,\lambda_2,\lambda_3)=a \spa{1}.{2}^b\spa{2}.{3}^c\spa{1}.{3}^d, \nn \\
A_3^{(-)} \equiv A_3(\tlambda_1,\tlambda_2,\tlambda_3)=a \spb{1}.{2}^b\spb{2}.{3}^c\spb{1}.{3}^d,
\label{ThreeVertexPlus}
\eea
where $a,b,c$ and $d$ are numbers which may be determined by calculation ($b,c,d$ alone may be determined by Lorentz symmetry~\cite{BenincasaCachazo}). For example, the pure-gluon amplitudes, where leg 1 has opposite helicity to that of leg 2 and 3, are
\bea
A_3^{(+),\tree}(1^+,2^-,3^-)=i\frac{ \spa{2}.{3}^3}{ \spa{1}.{2} \spa{3}.{1}} \,,\nn \\
A_3^{(-),\tree}(1^-,2^+,3^+)=-i\frac{ \spb{2}.{3}^3}{ \spb{1}.{2} \spb{3}.{1}}\,.
\label{gluonThreeVertex}
\eea
Note that the $A_3^{(+),\tree}(1^+,2^-,3^-)$ amplitude vanishes for the kinematics of a \MinusVertex\ vertex, as the numerator vanishes faster than the denominator $\sim 0^3/0^2$. Similarly, the $A_3^{(-),\tree}(1^-,2^+,3^+)$ amplitude vanishes on the \PlusVertex\ vertex. The same is true for all three-particle amplitudes in massless gauge and gravity theories, as these amplitudes are completely local objects (the spinor--helicity formalism obscures this fact).

As we are interested in cuts that involve more than one on-shell three vertex, we should also understand the properties of configurations of several such vertices.  One immediate simple ``transitivity'' property is: if two three-vertices of the same chirality are connected by a line with momentum $\lambda \tlambda$, then the two vertices are supported on the same chiral branch; thus, all momenta of the two vertices are proportional to a single spinor: $\lambda $ or $\tlambda$ depending on the chirality. As illustrated in \fig{ChiralProductFigure}, this implies that one can treat the kinematic configuration as an effective chiral four-point vertex. Similarly, if one encounters a connected chain of three-point vertices of the same chirality, one may treat it as an effective chiral higher-multiplicity vertex. All intermediate channels of such an effective kinematic vertex are on-shell, as all legs are proportional to the same spinor (note that there exists no known corresponding effective amplitude, the vertex here is used only as  a kinematic tool). 

%%%%%%%%%%%%%%%%%%%%
%FIGURE
\begin{figure}
\centerline{\epsfxsize 4 truein \epsfbox{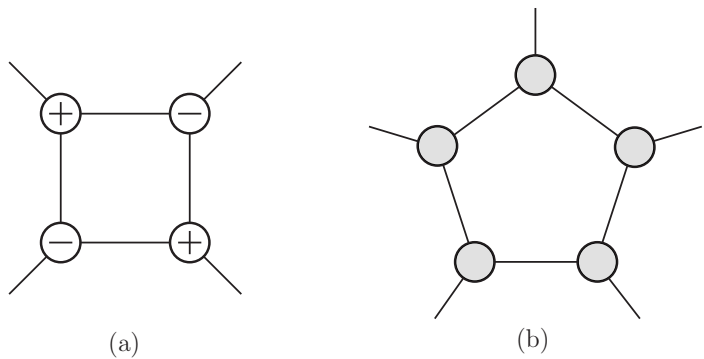}} \caption{A quadruple cut of a four-point amplitude in $D=4$, and a penta-cut of a five-point amplitude in $D>4$. In $D=4$, we use \PlusVertex\ and \MinusVertex\ labels on three-point vertices to specify the chirality, but in $D>4$ the chirality is not a well-defined notion.} \label{QuadrupleCutFigure}
\end{figure}
%%%%%%%%%%%%%%%%%%%%%%

\begin{figure}
\centerline{\epsfxsize 5 truein \epsfbox{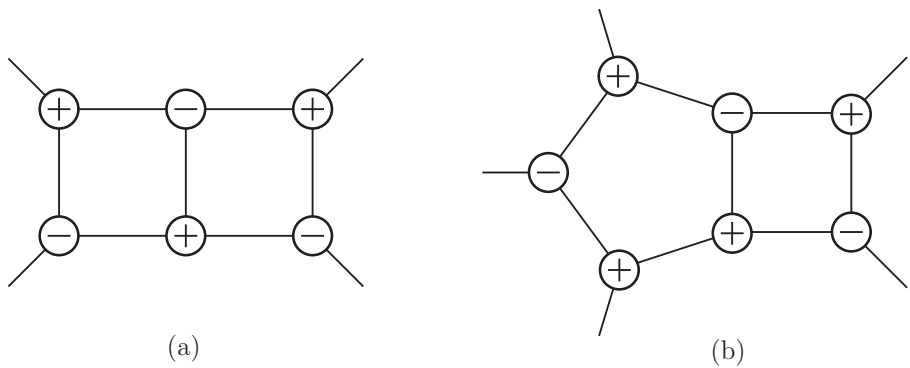}} \caption{A hepta-cut of a four-point amplitude, and an octa-cut of a five-point amplitude, both in $D=4$.} \label{PentaBoxFigure}
\end{figure}

\begin{figure}
\centerline{\epsfxsize 4.6 truein \epsfbox{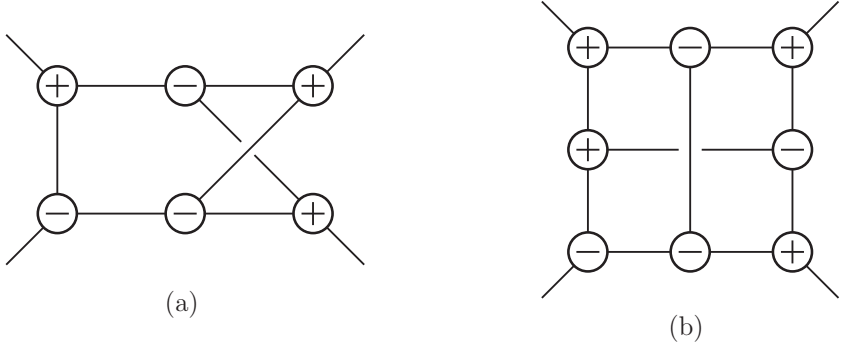}} \caption{A non-planar hepta-cut of a two-loop  four-point amplitude, and a non-planar deca-cut of a three-loop amplitude, both in $D=4$.} \label{non-planarFigure}
\end{figure}

In figures \ref{QuadrupleCutFigure}, \ref{PentaBoxFigure} and \ref{non-planarFigure} we illustrate typical cuts that are built out of only on-shell three point vertices. The four-particle quadruple cut in \fig{QuadrupleCutFigure}(a) uses four-dimensional momenta and must be built with vertices of alternating chirality~\cite{GeneralizedUnitarity2,BCF}. (The non-alternating configurations (not illustrated) correspond to degenerate situations where the momenta of two external legs are ``half-collinear''; that is, they have a common spinor, similar to the situation in \fig{ChiralProductFigure}.)
In $D>4$, the penta-cut in \fig{QuadrupleCutFigure}(b) is well defined, but unlike the $D=4$ case, the three vertices no longer come with definite chirality; therefore, we simply mark the vertices by a light shade (to avoid confusing them with tiny loops).
In \fig{PentaBoxFigure}, we show examples of two-loop cuts in $D=4$. Because the hepta-cut (a) enforces seven on-shell propagators $l_i^2=0$, one out of the eight degrees of freedom of the loop momenta is unconstrained (see discussion below). For the $D=4$ octa-cut of the five-point amplitude, \fig{PentaBoxFigure}(b), the number of on-shell conditions matches that of the number of loop momentum parameters; thus, the loop momenta is frozen by the cut, similar to the situation of the one-loop quadruple cut.

%%%%%%%%%%%%%%%%%%%%
%FIGURE
%%%%%%%%%%%%%%%%%%%%%%

%%%%%%%%%%%%%%%%%%%%
%FIGURE
%%%%%%%%%%%%%%%%%%%%%%

Before we conclude this section by working out a concrete example of a \NeqFoursYM\ maximal cut, it is convenient to recall some simplifying properties of state sums in maximal cuts~\cite{FiveLoop}. If a gluon of helicity $(+1)$ is attached to a chiral $(+)$ vertex, then the corresponding fermion and scalar amplitudes vanish, only the pure gluon amplitude has support:
\bea
A_3^{(+),\tree}(1_g^+,2_f,3_f)=A_3^{(+),\tree}(1_g^+,2_s,3_s)=0\,, \nn \\
A_3^{(+),\tree}(1_g^+,2_g^{-},3_g^{-})\neq 0\, .
\label{vanishing}
\eea
The same is true for a gluon of helicity $(-1)$ that is attached to a chiral $(-)$ vertex. It is easy to understand these vanishings from the point of view of charge conservation. For a massless fermion the helicity has to be conserved in gluon--fermion interactions; thus, in our all out-going notation the helicity must flip. For example,  $A_3^{(+),\tree}(1_g^+,2^-_f,3^-_f)$ is forbidden by helicity conservation. The case $A_3^{(+),\tree}(1_g^+,2^+_f,3^-_f)$ is allowed by helicity conservation, but it vanishes by virtue of having the wrong chirality. Similarly for a complex scalar its $U(1)$ charge has to be conserved in gluon--scalar interactions, implying that only amplitudes of the form $A_3^{(+),\tree}(1_g^+,2^+_s,3^-_s)$  are valid (where in this case $\pm$ ``scalar helicity'' refers to the scalar and its complex conjugate), but again this amplitude vanishes because of the chirality mismatch. The only nonvanishing amplitude is thus $A_3^{(+),\tree}(1_g^+,2^-_g,3^-_g)$, which has the right chirality. This ``projection property'' allows for some great simplifications of the maximal cuts by fine tuning the external states, such that all, or almost all, of the complicated particle and state sums in the loops drop out. In fact, \eref{vanishing} is generic to any massless gauge theory (and gravity); however, the most useful application is to the \NeqFoursYM\ theory where the amplitudes assemble into superamplitudes that are relatively insensitive to the choice of external states.

\begin{figure}
\centerline{\epsfxsize 5.5 truein \epsfbox{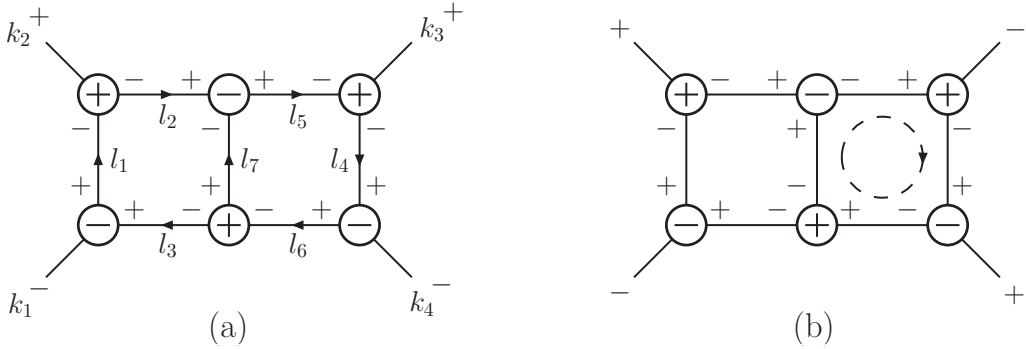}}
\caption{A singlet hepta-cut (a) and one of the two helicity
configurations (b) in a non-singlet cut.  The latter allows
gluons, fermions and scalars to propagate in the loop indicated by a
dashed circle.  The other configuration is obtained by flipping all
the helicity signs of the lines in this loop.  All lines in this figure
are cut and carry on-shell momenta. The arrows in (a) refer to the direction
of momentum flow.}
\label{singlet}
\end{figure}

If we classify the various state-sum cut contributions according to the helicity configuration of the particles (thus ignoring the particle species labels), then the greatest simplification that can arise in a nontrivial kinematic solution is the restriction to a single allowed helicity configuration
for the internal lines. We will refer to this configuration as a ``singlet''.  In this case only gluons can propagate inside the
diagram, as in \fig{singlet}(a).  Fermions or scalars are not allowed because in each loop there is at least one vertex of the wrong
chirality type.  The second-strongest simplification allows two helicity configurations. In such configurations, the particle content is purely gluonic except for one loop in which any particle type can propagate, as shown in
\fig{singlet}(b). (A single fermionic loop always allows two helicity
assignments, corresponding to interchanging fermion and antifermion,
and the same is true for a loop of complex scalars.)

%%%%%%%%%%%%%%%%%%%%
%FIGURE
%%%%%%%%%%%%%%%%%%%%%%
%

Now we continue with an explicit example of a maximal cut with only cubic vertices. Consider the two-loop four-point  hepta-cut in the singlet configuration shown in \fig{singlet}(a).  Although for analytical computations is not strictly necessary to solve the on-shell cut conditions, for illustrative purposes, we do so here. The seven on-shell conditions $l_i^2=0$, together with
momentum conservation at the six cubic vertex give a set of equations with the solution
\be
\begin{array}{llll}
l_1=\xi \lambda_1 \tlambda_2 \,, & l_2=l_1-k_2\,, &l_3=l_1+k_1\,,  \\
l_4=  \lambda_4 \tlambda_3 \frac{\vphantom{\tilde A}\spa{3}.{l_2}}{\spa{l_2}.{4}}\,, & l_5=k_3+l_4\,,& l_6=l_4-k_4\,,  & l_7=l_5-l_2\,,\\
\end{array} \label{ksolution}
\ee
where $\xi$ is an arbitrary complex parameter, corresponding to the
remaining degree of freedom in the integration not frozen by the
hepta-cut. The procedure to obtain this solution is quite simple. First we note that $l_1$ has to be proportional to the external spinors of the vertices that it connects to, namely, $l_1\propto \lambda_1 \tlambda_2$. We may try to find the overall scale of this momentum, but we observe  that there is  one  degree of freedom unfixed by this cut, therefore, we may account for this by introducing an arbitrary complex scale $\xi$ as in \eqn{ksolution}. Thus, we consider $l_1$ solved. Next we note that both $l_2$ and $l_3$ are on shell simply by momentum conservation. Therefore, we  proceed by solving $l_4$. Similar to $l_1$ it has to take the form $l_4= \xi' \lambda_4 \tlambda_3 $. Again by momentum conservation, the neighboring momenta $l_5$ and $l_6$ become on-shell using this ansatz. However, the momentum $l_7$ is off shell for generic values of $\xi$ and $\xi'$. Indeed the condition $0=l_7^2=(k_3+l_4-l_2)^2$ is the equation that allows us to solve for one of these parameters, say $\xi'$. The reader may check that this equation is linear in  $\xi'$ and that the solution takes the form given in  \eqn{ksolution}.

The hepta-cut of \fig{singlet}(a), for external gluons, is given by
\bea
A_4^{(2)}\Big|_{\rm 7\hbox{-}cut}&=&i^7 A^{\tag1}_{(1)} A^{\tag2}_{(2)}A^{\tag3}_{(3)}A^{\tag4}_{(4)}
A^{\tag5}_{(5)}A^{\tag6}_{(6)}\nn \\
&=&-i \frac{\spb{l_1}.{l_3}^3}{\spb{l_1}.{1}\spb{1}.{l_3}}   \frac{\spa{l_1}.{l_2}^3}{\spa{l_1}.{2}\spa{2}.{l_2}} \frac{\spa{l_5}.{l_4}^3}{\spa{l_5}.{3}\spa{3}.{l_4}} \nn \\ && \times \frac{\spb{l_6}.{l_4}^3}{\spb{l_6}.{4}\spb{4}.{l_4}} \frac{\spb{l_2}.{l_5}^3}{\spb{l_2}.{l_7}\spb{l_7}.{l_5}} \frac{\spa{l_3}.{l_6}^3}{\spa{l_3}.{l_7}\spa{l_7}.{l_6}}   \nn \\
&=& -i s^2 t \frac{\spa{1}.{4}^4}{\spa{1}.{2}\spa{2}.{3}\spa{3}.{4}\spa{4}.{1}} = -s^2tA^\tree_4(1^-,2^+,3^+,4^-) \,,
\label{singletstatement}
\eea
where $A^{\tag1}_{(i)}$ schematically denotes the three-point subamplitudes of \fig{singlet}(a), and $s=(k_1+k_2)^2$,  $t=(k_2+k_3)^2$ are the familiar Mandelstam variables. The intermediate steps in this calculation can be checked numerically using the solution \eref{ksolution}, or by simply massaging the analytical expressions using basic properties of spinor products. (For example, in order to simplify seemingly complicated products of tree amplitudes, it may be useful to keep in mind that maximal cuts are always local objects in the loop momenta.) The reader may confirm that \eqn{singletstatement} holds for any (nonzero)
value of the arbitrary parameter $\xi$.  

As  \eqn{singletstatement} is a unitarity cut in the $\NeqFour$ theory, we should have summed over all particle states as well as helicities in the loops; however, as argued this particular kinematic configuration, with conveniently chosen external helicities, is a singlet cut. There is only a single purely gluonic  contribution to the state sum! For a complete treatment of the two-loop four-point hepta-cut, one should also consider other kinematic solutions corresponding to other choices of chirality configurations for the cubic vertices. In general, this will result in non-singlet cuts, thus requiring the full supersymmetric multiplet of states. For such cases, it is useful to employ the four-dimensional on-shell superspace formalism of~\cite{Nair,GGK,FreedmanGenerating,DualSuperconformalSymmetry,KorchemskyOneLoop,AHCKGravity}; for additional details of on-shell superspace in multiloop cuts, see~\cite{SuperSum}. However, for the purpose of this review, it is sufficient to note that the cuts of the other kinematic solutions give the same result as \eref{singletstatement}, as shown by \cite{BuchbinderCachazo}.

\begin{figure}[t]
\centerline{\epsfxsize 5 truein \epsfbox{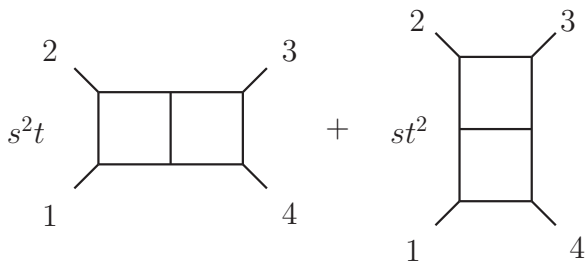}} \caption{An ansatz for the planar  four-point two-loop \NeqFoursYM\  amplitude in terms of the scalar integrals $I^{\rm (P)}(s,t)$ and $I^{\rm (P)}(t,s)$, as given by the computation of the maximal cuts. An overall factor of $-A^\tree_4(1,2,3,4)$ is suppressed. Other cuts, such as near maximal cuts, are needed to detect any terms missing from this ansatz. (However, no corrections are to be found as indeed this is the full planar amplitude~\cite{BRY}.) } \label{twoloopFigure}
\end{figure}

We have now computed the hepta-cut in one of the possible channels of the planar two-loop four-point amplitude; this allows us to write up an ansatz for the Feynman-like diagram in this ``horizontal'' channel: a double box scalar integral $I^{\rm (P)}(s,t)$ times the coefficient we computed in the cut $-s^2tA^\tree_4(1^-,2^+,3^+,4^-)$ . There is a second possible channel corresponding a ``vertical'' double box integral  $I^{\rm (P)}(t,s)$, see \fig{twoloopFigure}. The hepta-cut corresponding to this diagram can be computed the same way as the first hepta-cut \eref{singletstatement}. The result, which we will simply quote, is remarkably similar to \eref{singletstatement}, the only difference is a exchange $s \leftrightarrow t$ (note that the helicity amplitude $A^{(2)}_4(1^-,2^+,3^+,4^-)$ has no cyclic symmetry or flip symmetry that automatically enforces this). Finally, with the results of the maximal cuts in our hands we can write down an ansatz for the full planar amplitude:
\be
A^{(2)}_4(1,2,3,4)=-A_{4}^{\tree}(1,2,3,4)\left(s^2 t I^{\rm (P)}(s,t)+s t^2 I^{\rm (P)}(t,s)\right)\,,
\label{2loopamp}
\ee
where we have suppressed the helicity assignment $(1^-,2^+,3^+,4^-)$ that we have been considering,  the coupling constant, and the usual planar color structure of adjoint particles.  If, rather, we were to compute the maximal cuts of the remaining four point amplitudes, for other gluon helicity assignments $(1^\pm,2^\pm,3^\pm,4^\pm)$, or for the external fermions and scalars, the result is the same as \eref{2loopamp} (this fact can be attributed to special properties of the four-point function of \NeqFoursYM\ which hold in all dimensions $D\le10$, see \sect{RecursiveCutsSection}).

%%%%%%%%%%%%%%%%%%%%
%FIGURE
%%%%%%%%%%%%%%%%%%%%%%

\subsection{On-shell three-vertices in higher dimensions.} 
In $D>4$, the structure of a single three-point vertex is quite similar to the $D=4$ case, as the three momenta $p$, $q$, and $r$ can be embedded in a four-dimensional subspace (see \eref{embededPQ}) where the details are the same as above. However, the two-fold branching is no longer a meaningful notion in $D>4$ as one may smoothly connect the two chiral solutions of a given subspace by a rotation into the dimensions orthogonal to the subspace. Also the simple $D=4$ forms of the amplitudes cannot be used in four-dimensional subspaces, as the $D=4$ spinors (and amplitudes) transform nontrivially under the full $D$-dimensional Lorentz rotations.  Instead, for the $D$-dimensional vertices one may simply use the fully covariant form of the amplitude, {\it e.g.} for three gluons,
\be
A_3^{\tree}(1,2,3)=\varepsilon_1\cdot \varepsilon_2 ~ \varepsilon_3^\mu (k_1-k_2)_\mu~+~{\rm cyclic}\,,
\label{CovariantThreeVertex}
\ee
which is simply the familiar Feynman gauge vertex contracted with the gluon polarizations.
Alternatively one may find useful the explicit generalizations of the spinor--helicity notation in $D=6$ \cite{Donal,SixDimSusy,SixDim} and in $D=10$ \cite{Donal10}.

Since much of the properties of the $D=4$ three-point vertex carries over to higher dimensions, it should also be true that there are particular choices of gluon polarizations that will remove the contributions of the fermion and scalar amplitudes. Indeed, choosing the polarization vector of the gluon to be proportional to the momentum of one of the neighboring lines is the correct $D$-dimensional generalization of \eref{vanishing},
\bea
A_3^{\tree}(1_g,2_f,3_f)=A_3^{\tree}(1_g,2_s,3_s)=0\,, \hskip 1.5cm (\varepsilon_1^{\mu}\propto k_2^{\mu}) \nn \\
A_3^{\tree}(1_g,2_g,3_g)\neq 0\,.
\eea
This is easily seen to be true from the fermion vertex $\overline{\psi}_3 \varepsilon_1^\mu \gamma_\mu \psi_2$, because of the Dirac equation $k_2^\mu\gamma_\mu \psi_2=0$ , and from the scalar vertex $\varepsilon_1^\mu(k_2-k_3)_\mu$, because $k_i \cdot k_j=0$ for three-point kinematics.

It should be noted that in some cases it is convenient to carry out the method of maximal cuts (or other schemes of generalized unitarity) in four dimensions where the amplitudes entering the cuts are quite simple and, after arriving at a well-behaved ansatz, verify completeness using higher-dimensional spanning cuts.  In particular, for four-particle amplitudes in  the \NeqFoursYM\ theory, this strategy has been successfully carried out through four loops~\cite{CompactThree,Neq44np}.

\subsection{Cuts with higher-multiplicity vertices}

\begin{figure}
\centerline{\epsfxsize 4.5 truein \epsfbox{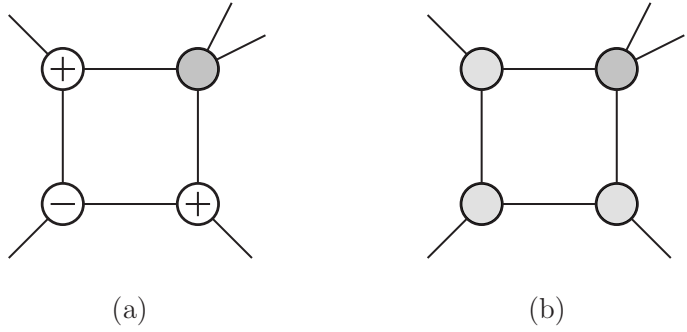}} \caption{A quadruple cut in $D=4$ is a maximal cut, whereas a quadruple of the same five-point amplitude in $D>4$ is a near-maximal cut, or more precisely a next-to-maximal cut.} \label{NearMaxFigure}
\end{figure}

Maximal cuts do not only include generalized cuts with three-point vertices; according to the definition \eref{MCdef}, higher multiplicity vertices are allowed. This situation occurs, for example, in the five-point one-loop quadruple cut in four dimensions, displayed in \fig{NearMaxFigure}(a). The four-point vertex may contain a propagator corresponding to a pentagon integral, but the constraints of four-dimensional spacetime prohibits all five propagators to be on-shell simultaneously (for non-degenerate external momenta).  The number of cut conditions is saturated by the one-loop quadruple cut in four dimensions, and thus quadruple cuts are considered maximal in this spacetime. 
However, for $D>4$, the five-point one-loop amplitude evaluated on the penta-cut exists and is considered a maximal cut, and the quadruple cut, displayed in \fig{NearMaxFigure}(b), would be considered a next-to-maximal cut, as per the discussion in \sect{maxcutdefSection}. 

%%%%%%%%%%%%%%%%%%%%
%FIGURE
%%%%%%%%%%%%%%%%%%%%%%

From a technical point of view, the process of carrying out maximal cuts and near-maximal cuts that involve higher-multiplicity subamplitudes is quite similar to the case of maximal cuts involving only three-point vertices.  The main distinction is that higher-point vertices have no definite chirality, not even in four dimensions. This implies that the phase space of a four dimensional cut with many higher point vertices is in general smoother than four-dimensional cuts with only three-point vertices.  The chiral branching of the four dimensional kinematics entering a cut is entirely due to the presence of four-dimensional on-shell thee-point vertices.

\begin{figure}
\centerline{\epsfxsize 6 truein \epsfbox{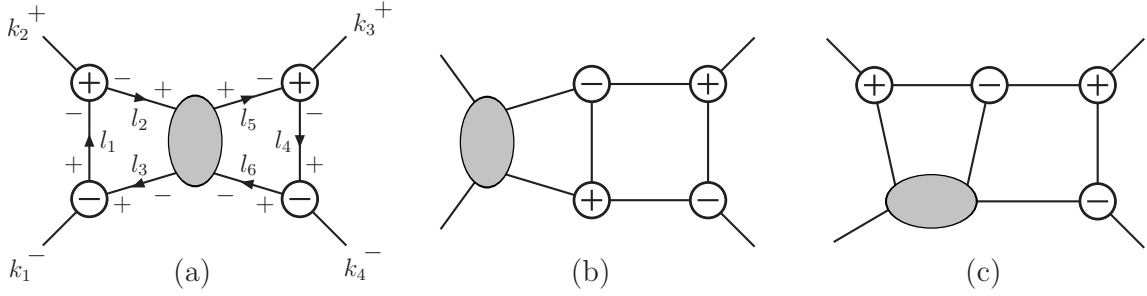}} \caption{ Planar next-to-maximal cuts of the four-point two-loop amplitude. } \label{nearmaximalFigure}
\end{figure}

For an example of a near-maximal cut with a four-point subamplitude, we will consider the two-loop four-point amplitude again. We wish to show that the ansatz \eref{2loopamp} receives no more corrections at the level of the next-to-maximal hexa-cuts. First we list the different cut topologies that can arise at the next-to-maximal level. Starting from the maximal cut topology in \fig{singlet}, we may relax the cut condition of any one of the on-shell lines. This gives the three cut topologies in shown in \fig{nearmaximalFigure}. As before, these cuts have to be considered for all possible kinematic solutions (assignment of the vertex chirality) and for all possible channels (cyclic and anticyclic permutations of the external states). We will work out the \fig{nearmaximalFigure}(a) hexa-cut explicitly. Note that the vertex chirality and labeling is the same as for \fig{singlet}(a) except that momentum $l_7$ is no longer on-shell. We may recycle the kinematic solution \eref{ksolution}, with the only difference that $\xi'$ is now arbitrary, and thus $l_7$ is off-shell:
\be
\begin{array}{llll}
l_1=\xi \lambda_1 \tlambda_2 \,, & l_2=l_1-k_2\,, &l_3=l_1+k_1\,,  \\
l_4= \xi'  \lambda_4 \tlambda_3 \,, & l_5=k_3+l_4\,,& l_6=l_4-k_4\,,  & l_7=l_5-l_2\,.
\end{array} \label{ksolution2}
\ee
The hexa-cut is given by
\bea
A_4^{(2)}\Big|_{\rm 6\hbox{-}cut}\hskip-2mm&=&
i^6 A^{\tag1}_{(1)} A^{\tag2}_{(2)}A^{\tag3}_{(3)}A^{\tag4}_{(4)}
A^{\tag5}_{(5')}\nn \\&=&
-i \frac{\spb{l_1}.{l_3}^3}{\spb{l_1}.{1}\spb{1}.{l_3}}   \frac{\spa{l_1}.{l_2}^3}{\spa{l_1}.{2}\spa{2}.{l_2}}\frac{\spa{l_5}.{l_4}^3}{\spa{l_5}.{3}\spa{3}.{l_4}}   \frac{\spb{l_6}.{l_4}^3}{\spb{l_6}.{4}\spb{4}.{l_4}}   \frac{\spa{l_3}.{l_6}^3}{\spa{l_3}.{l_2}\spa{l_2}.{l_5}\spa{l_5}.{l_6}}  \nn \\
&=& -i s^2 t  \frac{1}{l_7^2} \frac{\spa{1}.{4}^4}{\spa{1}.{2}\spa{2}.{3}\spa{3}.{4}\spa{4}.{1}}\nn \\
&=&  -s^2t \frac{1}{l_7^2} A^\tree_4(1^-,2^+,3^+,4^-) \,,
\label{nexttomax}
\eea
where again $A^{\tag1}_{(i)}$ schematically denotes the three- and four-point subamplitudes of the cut, in this case \fig{nearmaximalFigure}(a). The result is surprisingly similar to the hepta-cut \eref{singletstatement}, the only difference is the appearance of the propagator $1/l_7^2$. This a positive result showing that our ansatz for the amplitude passed the test. Indeed the reader may check that the same hexa-cut on the scalar double box integral in our convention \eref{cutdef2} is
\be
I^{\rm (P)}(s,t)\Big|_{\rm 6-cut} = \frac{1}{l_7^2} \,,
\ee
explaining the appearance of this propagator factor. Finally, we note that evaluating all the hexa-cuts of \fig{nearmaximalFigure} for the various kinematic solutions and channels, informs us that the ansatz \eref{2loopamp} receives no more corrections at the next-to-maximal cut level. This is a strong indication that the ansatz for the amplitude may be complete.  To ensure that the ansatz is correct, one would have to evaluate it on a complete set of cuts in $D$ dimensions, such as the two- and three-particle unitarity cuts (done in ref.~\cite{BRY} in $D=4$). This we will not do explicitly here (although the next section discusses the two-particle cut); we simply note that \eref{2loopamp} is indeed the full planar amplitude~\cite{BRY}.

%%%%%%%%%%%%%%%%%%%%
%FIGURE
%%%%%%%%%%%%%%%%%%%%%%

\section{Some recursive $D$-dimensional cuts}
\label{RecursiveCutsSection}

In the previous section, we discussed unitarity cuts which rely on cutting all or almost all propagators of every internal loop.  As a contrast, here we will consider cuts where many of the propagators are left uncut.   Specifically, will review two types of very useful cuts, particular to four-point \NeqFoursYM, namely the well-known ``two-particle cut'' of the four-point amplitude~\cite{BRY,BDDPR}, and the more recent ``box cut''~\cite{Neq44np};  both cuts can be expressed very simply, and in all generality, in $D$ dimensions, in terms of lower-loop expressions. These cuts are useful for computing contributions to the $D=4-2\epsilon$ dimensionally regulated amplitudes; in this case the regularization scheme implied is the four-dimensional-helicity (FDH) version of dimensional reduction~\cite{DimensionalReductionScheme}. And similarly, for amplitudes in $D\le 10$, the corresponding theory implied is the dimensional reduction of the ten dimensional ${\cal N}=1$ sYM theory.  

Before introducing the two cuts, we will review a structure present in the four-point amplitudes of \NeqFoursYM, which plays an important role in the subsequent discussion, due to the especially simple dependence on the external states.  It is known that all \NeqFoursYM\ four-point amplitudes can be expressed in a common factorized form~\cite{Neq44np}
\be
{\cal A}_4^{(L)}(1,2,3,4)
= g^{2L + 2} {\cal K}(1,2,3,4) \, \UniversalFactor^{(L)}(1,2,3,4)\,,
\label{FourPointFactorization}
\ee
where ${\cal A}_4$ represents the full color-dressed $D\le10$-dimensional amplitude ({\it cf.} the color-stripped partial amplitude $A_4$), the kinematic prefactor ${\cal K}$ is the same in all loop-diagrams, and the loop-specific factor $U$ is independent of the states of the external particles.  The $D$-dimensional  \NeqFoursYM\ amplitudes are defined as amplitudes of the ten-dimensional ${\cal N}=1$ sYM theory dimensionally reduced to $D$ dimensions.
All of the $\NeqFour$ state dependence in ${\cal A}_4$ ($16^4$ configurations of states) is carried by the kinematic prefactor 
\be
{\cal K}(1,2,3,4) \equiv s_{12} \, s_{23} \,A_4^{\tree}(1,2,3,4)\,,
\label{Kprefactor}
\ee
where $A_4^{\tree}(1,2,3,4)$ is the canonically color-ordered $D$-dimensional tree amplitude. Note that multiplying the tree by $s_{12} s_{23}=(k_1+k_2)^2(k_2+k_3)^2$ renders the resulting expression fully crossing symmetric; hence, the ordering of  the arguments of ${ \cal K}$ indicates no specific color order. 

In four dimensions, a compact form of ${ \cal K}$ can be written down using anti-commuting parameters $\eta$ in
the four-dimensional on-shell superspace formalism~\cite{Nair,GGK,FreedmanGenerating,DualSuperconformalSymmetry,KorchemskyOneLoop,AHCKGravity},
\bea
A_{4}^{\tree}(1,2,3,4)\Bigr|_{D=4} &=&
{i\delta^{(8)}(\lambda_1 \eta_1+\lambda_2 \eta_2
              +\lambda_3 \eta_3+\lambda_4 \eta_4)
 \over \spa{1}.{2}\spa{2}.{3}\spa{3}.{4}\spa{4}.{1}} \,,
\label{SuperAmplitude} \\
{\cal K}(1,2,3,4)\Bigr|_{D=4} &=&
-i \, {\spb1.2\spb3.4 \over \spa1.2\spa3.4}
\, \delta^{(8)}(\lambda_1 \eta_1+\lambda_2 \eta_2
              +\lambda_3 \eta_3+\lambda_4 \eta_4) \,.
\label{SuperAmplitudeII}
\eea
It is not difficult to verify that $(\spb1.2\spb3.4)/(\spa1.2\spa3.4)$
is symmetric under exchange of any two legs. Similar compact superspace expressions for the four-point amplitude exist in $D=6$~\cite{SixDimSusy} using generalized spinor-helicity notation~\cite{Donal}, and  in $D=10$~\cite{Mafra} using Berkovits' pure spinor formalism~\cite{Berkovits}.
Related to this, ${\cal K}$ represents the color-stripped four-point (linearized)
matrix elements of the local operator $\Tr F^4$, plus its
supersymmetric partners.  Therefore, ${\cal K}$ is a natural prefactor
to extract from the four-point amplitude in \NeqFoursYM.

All of the color dependence of the amplitude is carried by the external-state-independent ``universal factor'' $\UniversalFactor^{(L)}$.  This  factor at tree level, $\UniversalFactor^{(0)}$, is given by~\cite{Neq44np},
\be
{\UniversalFactor}^{(0)}(1,2,3,4) = 
 \biggl(
 \frac{\f^{a_4a_1b}\f^{ba_2a_3}} {s_{12} s_{23}}
+\frac{\f^{a_3a_1b}\f^{ba_2a_4}} {s_{12} s_{13}}
 \biggr)\,.
\label{TreeUniversalFactor1}
\ee
In general, the universal factor $\UniversalFactor^{(L)}$ is a sum of $L$-loop integrals.  The integrands entering $\UniversalFactor^{(L)}$ are rational functions of momentum invariants involving the loop and external momenta.  The universal factors for $L=1,2,3,4$, including planar and non-planar contributions, may be found by in~\cite{Neq44np}, and planar $L=5$ in~\cite{FiveLoop}. 

The four-point factorization \eref{FourPointFactorization} makes it possible to immediately write down simple $D$-dimensional expression for  cuts involving this amplitude, usually in terms of lower-loop numerator factors.  We will now review two related special classes of cuts, which owe their simple structure to this property. These cuts  can be applied to both planar and non-planar amplitudes, making them especially powerful.

%%%%%%%%%%%%%%%%%%%%%%%%%%%%%%%%%%%%

\subsection{Two-particle cuts of the four-point \NeqFoursYM\ amplitude}
\label{TwoParticleCutSection}

%%%%%%%%% FIGURE %%%%%%%%%%%%%%%
\begin{figure}[tbh]
\centerline{\epsfxsize 1.9 truein \epsfbox{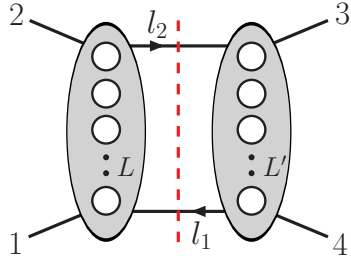}}
\caption[a]{\small A two-particle cut that may be used to construct
contributions to the $(L+L'+1)$-loop amplitude from those at
$L$ and $L'$ loop orders.}
\label{TwoParticleCutFigure}
\end{figure}
%%%%%%%%%%%%%%%%%%%%%%%%%%%%%%%%

A regular {\it two-particle cut} of a multiloop four-point amplitude has the form shown in \fig{TwoParticleCutFigure} --- it cuts the amplitude into two lower-loop four-point amplitudes. Using the factorization \eref{FourPointFactorization}, it will be possible to write down the cut in terms of lower-loop universal factors, thus straightforwardly determining a class of integrals in the amplitude that are detectable in this cut.  The result of this exercise
will be valid in $D$ dimensions, whenever the lower-loop universal factors are valid in $D$ dimensions.

We evaluate the generic two-particle color-dressed cut depicted in \fig{TwoParticleCutFigure}.  It cuts the $(L+L'+1)$-loop
amplitude ${\cal A}_4^{(L+L'+1)}(1,2,3,4)$ into the two four-point amplitudes ${\cal A}_4^{(L)}(-l_1,1,2,l_2)$ and ${\cal A}_4^{(L')}(-l_2,3,4,l_1)$ of loop orders $L$ and $L'$,
\be
{\cal A}_4^{(L+L'+1)}(1,2,3,4)\Big|_{\rm 2\hbox{-}cut}
\hskip-2mm = i^2 \, \sum_{\NeqFour \atop {\rm states}}
{\cal A}_4^{(L)}(-l_1,1,2,l_2) \, {\cal A}_4^{(L')}(-l_2,3,4,l_1) \,,
\label{GenLoopTwoParticleCut}
\ee
where the state sum is over the particles with momenta $l_1$ and $l_2$.

Using the factorization~(\ref{FourPointFactorization}) and the state-independence of $\UniversalFactor^{(L)}$, we can immediately rewrite the cut as follows:
\bea
 {\cal K}(1,2,3,4) \, \UniversalFactor^{(L+L'+1)}(1,2,3,4)\Big|_{\rm 2\hbox{-}cut}
\hskip-2.5mm &=& i^2 
\, \UniversalFactor^{(L)}(-l_1,1,2,l_2) \, 
  \UniversalFactor^{(L')}(-l_2,3,4,l_1) \nn\\
&& \hskip-1mm \times
  \sum_{\NeqFour \atop {\rm states}}
  {\cal K}(-l_1,1,2,l_2) \, {\cal K}(-l_2,3,4,l_1) \,. \nn\\
\label{GenLoopTwoParticleCutRewrite}
\eea
To evaluate this, we use the sewing relation between two four-point color-ordered \NeqFoursYM\ trees~\cite{BRY,BDDPR},
\be
\sum_{\NeqFour \atop \rm states}
\hskip-1mm  A_4^{(0)}(-l_1,1,2,l_2) A_4^{(0)}(-l_2,3,4,l_1)
=\hskip-1mm  - i \sv \tv \, A_4^{(0)}(1,2,3,4) {1\over s_{2l_2} s_{4l_1}} \,.
\label{SewingRelation}
\ee
This sewing relation is valid in any dimension $D$ and for any external states in the $\NeqFour$ multiplet.  A straightforward way to confirm \eqn{SewingRelation} is to work in $D=10$ and evaluate the sum over states in components, using the fact that in $D=10$ \NeqFoursYM\ is equivalent to an $\Neqone$ theory composed of a gluon and a gluino. By dimensional reduction the sewing relation~(\ref{SewingRelation}) then holds in any dimension $D\le 10$.  Recently, this equation has also been verified directly in six dimensions using an on-shell superspace~\cite{SixDimSusy}. 

Substituting in the relation $A_4^{(0)}={\cal K}/(s_{12}s_{23})$ given in \eref{FourPointFactorization}, we find that \eref{SewingRelation} is equivalent to
\bea
  \sum_{\NeqFour \atop {\rm states}}
  {\cal K}(-l_1,1,2,l_2) \, {\cal K}(-l_2,3,4,l_1)&=& -i\, s_{12}^2 \, {\cal K}(1,2,3,4)\,.
\label{KSewingRelation}
\eea

%%%%%%%%% FIGURE %%%%%%%%%%%%%%%
\begin{figure}[tb]
\centerline{\epsfxsize 5.8 truein \epsfbox{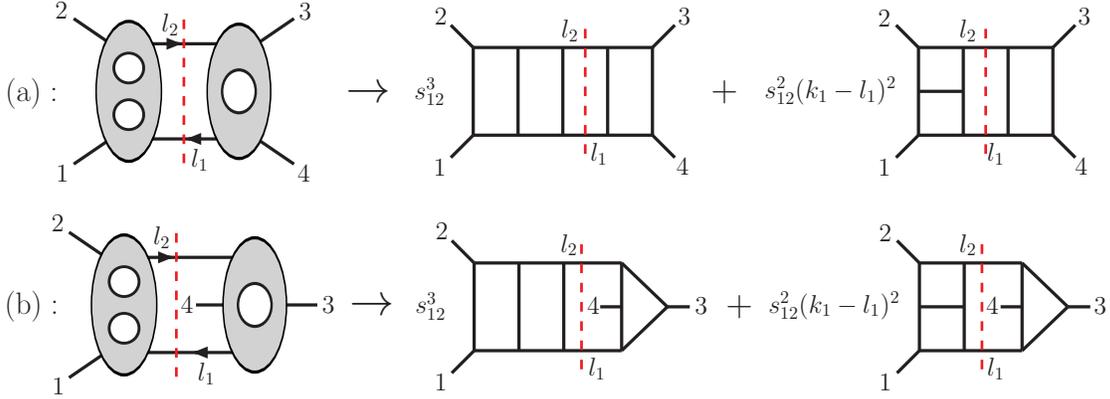}}
\caption[a]{\small Sample contributions to the full color-dressed
  two-particle cut for $L=2$ and $L'=1$.  The diagrams on the
  right-hand side show some of the terms in
  $\UniversalFactor^{(4)}(1,2,3,4)$ that are constructed from these
  cuts, using $\UniversalFactor^{(2)}$ and $\UniversalFactor^{(1)}$.  The explicit
  color factors, as well as factors of $i$, have been omitted.}
\label{TwoParticleCutConstructFigure}
\end{figure}
%%%%%%%%%%%%%%%%%%%%%%%%%%%%%%%%

Applying \eqn{KSewingRelation} to \eqn{GenLoopTwoParticleCutRewrite}, we find the key equation for building all contributions from two-particle cuts directly in terms of the $\UniversalFactor$s: 
\be
\UniversalFactor^{(L+L'+1)}(1,2,3,4)\Big|_{\rm 2\hbox{-}cut}
= i \, s^2_{12}\, \UniversalFactor^{(L)}(-l_1,1,2,l_2) 
             \, \UniversalFactor^{(L')}(-l_2,3,4,l_1) \,.
\label{TwoParticleUSewing}
\ee
\Eqn{TwoParticleUSewing} is rather powerful.  No complicated calculations remain in order to obtain all contributions visible in two-particle cuts; they are given simply by taking the product of  lower-loop results.  The color-dressed $\UniversalFactor^{(L+L'+1)}$ is given immediately as a sum over products of individual integrals residing inside the $\UniversalFactor^{(L)}$ and $\UniversalFactor^{(L')}$ factors, up to terms that vanish because of the on-shell conditions, $l_1^2 = l_2^2 = 0$.

\Fig{TwoParticleCutConstructFigure} illustrates diagrammatically some of the terms generated by \eqn{TwoParticleUSewing} for the case $L=2$ and $L'=1$. For simplicity, we draw only the planar contributions of $\UniversalFactor^{(2)}(-l_1,1,2,l_2)$, omitting the color factor $\tilde{f}^{abc}$s as well as all factors of $i$ (which can be restored at one's convenience).  The denominator factors in $\UniversalFactor^{(2)}$ and $\UniversalFactor^{(1)}$ correspond to the un-cut propagators that are graphically visible on the left- and right-hand sides of the dashed line, respectively.  Similarly, the numerator factor for each parent graph on the right-hand side of \fig{TwoParticleCutConstructFigure} is given by forming the product of the numerator factors for the one-loop box diagram and two-loop  double box (taking into account the proper permutation of legs), and then multiplying by two powers of $s_{12}$.

%%%%%%%%% FIGURE %%%%%%%%%%%%
\begin{figure}[t]
\centerline{\epsfxsize 3.3 truein \epsfbox{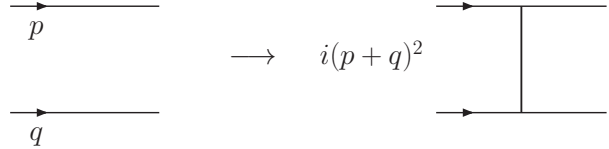}}
\caption[a]{\small The rung rule~\cite{BRY,BDDPR}
  for generating a higher-loop contribution by inserting vertical rung
between two lines inside a loop diagram of \NeqFoursYM\ amplitudes. The dimension of the extra propagator is compensated by the extra factor $i(p+q)^2$ in the numerator. This rule reproduces the numerators of planar diagrams generated by iterated two-particle cuts.}
\label{rungruleFigure}
\end{figure}
%%%%%%%%%%%%%%%%%%%%%%%%%%%%%

The simple formula~\eqn{TwoParticleUSewing} for the two-particle cuts provides a rather useful tool for generating many higher-loop contributions by recycling lower-loop ones.  Indeed, it is the mechanism behind the many of the diagrams given by the well-known ``rung rule''~\cite{BRY,BDDPR}, shown in \fig{rungruleFigure}. Unlike the two-particle cut formula, the rung rule is a heuristic rule that should be applied with some care; it can potentially suggest non-contributing terms when the two-particle cut does not apply~\cite{BCDKS}.  That warning aside, for planar \NeqFoursYM, the rung rule is rather powerful.  For example, it is quite interesting that  the rung rule generates terms that are invariant under the dual conformal symmetry of \NeqFoursYM~\cite{BCDKS,FiveLoop}, presaging modern awareness of this symmetry.

\subsection{Box cuts of multiparticle \NeqFoursYM\ amplitudes}
\label{BoxCutSection}

\begin{figure}[t]
\centerline{\epsfxsize 4.7 truein \epsfbox{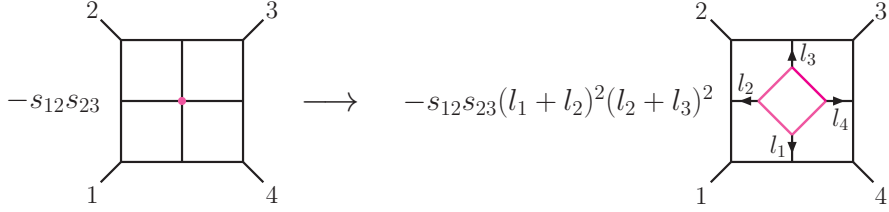}}
\caption[a]{\small The box-substitution rule~\cite{FiveLoop}
  for generating a higher-loop contribution by inserting a one-loop
  four-point box subintegral into a four-point vertex. In
  this example, we substitute a box into the central four-point vertex
  in the four-loop ``window'' diagram. The result is a five-loop
  integral that cannot be obtained from two-particle cuts or the rung rule. }
\label{BoxSubRuleFigure}
\end{figure}

The two-particle cut has a particularly nice generalization that also relies on the simple structure of the four-point amplitude of \NeqFoursYM.
This cut can be thought of as a (generalized) four-particle cut that isolates a four-point subamplitude, which in the simplest nontrivial case is a box subdiagram; thus, the cut was named  ``box cut'' in~\cite{Neq44np}. A crude version of this cut appeared already~\cite{FiveLoop} as a ``box-substitution rule''.  It allowed the
construction of $L$-loop contributions with a box subgraph, starting from $(L-1)$-loop contributions with a contact interaction, as illustrated in \fig{BoxSubRuleFigure}. Similar rules were discussed in conjunction with the leading singularity method in~\cite{CachazoSkinner}. Unlike the two-particle cut discussed in the previous section, the box cut is applicable to multiparticle amplitudes. Now we will review the derivation of the $D$-dimensional box cut.

%%%%%%%%% FIGURE %%%%%%%%%%%%
%%%%%%%%%%%%%%%%%%%%%%%%%%%%%

\begin{figure}[t]
\centerline{\epsfxsize 4.5 truein \epsfbox{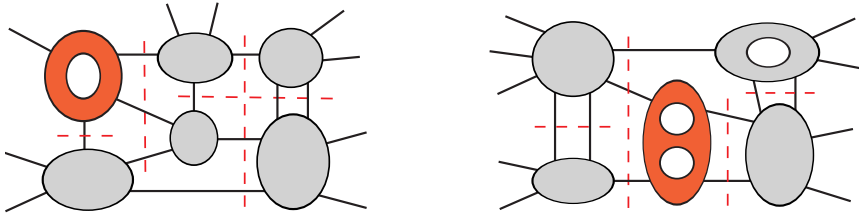}}
\caption[a]{\small Two examples of multiparticle multiloop ``box
  cuts''. They reduce to lower-loop cuts by replacing the red
  four-point subamplitudes by four-point color-ordered trees,
  multiplied by known numerator and denominator factors. 
  This property allows these cuts to be computed easily in $D$
  dimensions, once the amplitudes with fewer loops are
  known. White holes represent loops and the darker (red)
  subamplitudes mark four-point amplitudes amenable to reduction.}
\label{GeneralBoxCutFigure}
\end{figure}

Consider the generalized cut of an $L$-loop $n$-point amplitude,
\be
{\cal A}_n^{(L)}\Big|_{\rm box~cut} \equiv 
\sum_{\NeqFour \atop \rm states}
 {\cal A}_{(1)}\cdots {\cal A}_{4, (i)}^{(L')} \cdots {\cal A}_{(m)} \,,
\label{BoxCut}
\ee
that is composed of a generic set of color-dressed amplitude factors, except for the $i^{\rm th}$ such factor, which we take to be a color-dressed $L'$-loop four-point subamplitude, ${\cal A}_{4, (i)}^{(L')}$.   Example of such box cuts are given in \fig{GeneralBoxCutFigure}.

%%%%%%%%% FIGURE %%%%%%%%%%%%%%%%%%
%%%%%%%%%%%%%%%%%%%%%%%%%%%%%%%%

As mentioned, the $L'$-loop four-point subamplitude of \NeqFoursYM\ is special because of the factorization property
(\ref{FourPointFactorization}).  Labeling the cut momenta by $l_1,l_2,l_3,l_4$, we have
\be
 {\cal A}_{4,(i)}^{(L')}(l_1,l_2,l_3,l_4) 
= A_{4,(i)}^{(0)}(l_1,l_2,l_3,l_4)
\,  (l_1+l_2)^2 (l_2 + l_3)^2 
\, \UniversalFactor^{(L')}(l_1,l_2,l_3,l_4) \,, \,
\label{FourPointFactorizationBoxCut}
\ee
where, as in the previous section, we use ${\cal A}_4$ to represent the color-dressed amplitude, and only the color-ordered tree amplitude factor $A_{4,(i)}^{(0)}$ depends on the states crossing the cuts.  Therefore, we can pull the factor $\UniversalFactor^{(L')}$  out of the sum over states in \eqn{BoxCut}, leading to a simpler  expression in the summand:
\be
{\cal A}_n^{(L)}\Big|_{\rm box \atop cut}\hskip-1mm =(l_1+l_2)^2 (l_2 + l_3)^2
\UniversalFactor^{(L')}(l_1,l_2,l_3,l_4)
 \, \sum_{\NeqFour \atop \rm states} 
 {\cal A}_{(1)}\cdots  A_{4, (i)}^{(0)} \cdots  {\cal A}_{(m)} \,.\,
\label{BoxCutTrue}
\ee
The state-sum is identical to a lower-loop cut, that of the $(L-L')$-loop amplitude, but utilizing the color-ordered contribution to the $i^{\rm th}$ tree.  This fact immediately gives a simple relation between the $L$-loop box cut and contributions to the reduced $(L-L')$-loop cut under the same cut conditions.

We can formally write down an equation relating the cut of an $L$-loop amplitude to a cut of a lower-loop one as,
\be
{\cal A}_n^{(L)}\Big|_{\rm box~cut}
= (l_1+l_2)^2 (l_2+l_3)^2
\,\UniversalFactor^{(L')}(l_1,l_2,l_3,l_4) \, 
\tilde{{\cal A}}_n^{(L-L')}\Big|_{\rm cut} \,.
\label{BoxCutReduction}
\ee
We introduced the reduced cut $\tilde{{\cal A}}$ notation to emphasize that the state-sum in \eqn{BoxCutTrue} is exactly a $(L-L')$ loop unitarity cut which is color-dressed with $\f^{abc}$ everywhere, except for the four-point color-ordered tree amplitude whose associated color factors are accounted for in the $L'$-loop universal factor $\UniversalFactor^{(L')}$.

%%%%%%%%% FIGURE %%%%%%%%%%%%%%%%%%
\begin{figure}[t]
\centerline{\epsfxsize 5.5 truein \epsfbox{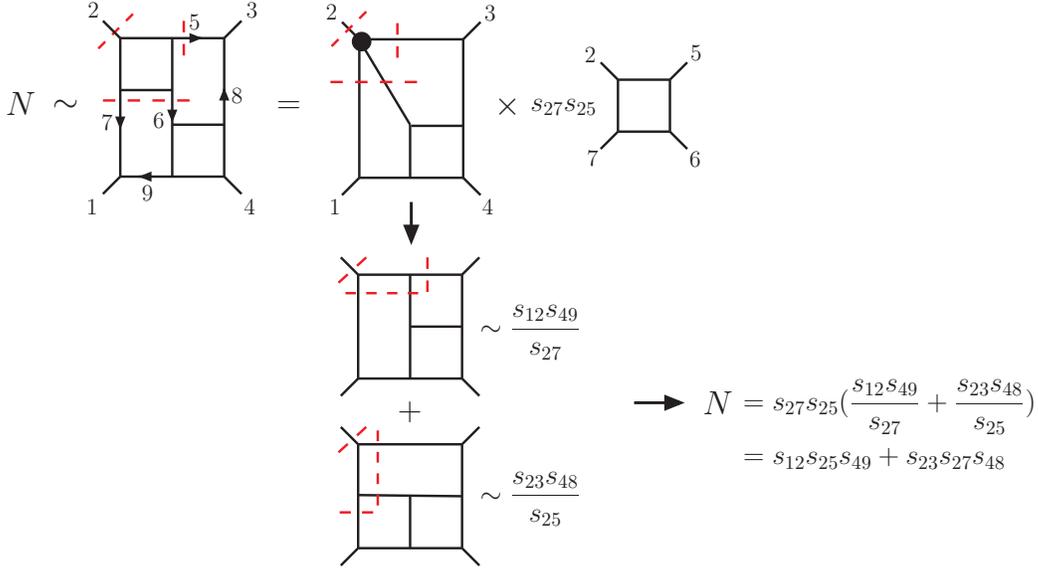}}
\caption[a]{\small Practical application of the box cut to determine the
  numerator $N$ for a four-loop integral.  The (red) dashed cut conditions around the
  upper left box in this diagram allow us to replace it by the product
  of a reduced cut diagram, some kinematic factors and a box
  integral.  The reduced cut diagram is then expanded into two
  three-loop ``tennis-court'' diagrams, corresponding to the two
  allowed channels of the marked four-point vertex.  The relevant
  kinematic pieces of the tennis-court diagrams, {\it i.e.} the
  numerators and the spurious propagators, are extracted from the
  known three-loop contribution. Assembling all the kinematic
  factors gives the result for $N$ (with the overall factor ${\cal K}$ suppressed), which is free of spurious
  propagators.}
\label{boxcutFigure}
\end{figure}
%%%%%%%%%%%%%%%%%%%%%%%%%%%%%%%%

\Fig{boxcutFigure} gives an example illustrating that the box cut is very a practical tool to work with. Using a back-of-the-envelope calculation, one can determine the numerator polynomial of a four-loop integral~\cite{BCDKS,Neq44np}, using the known three-loop four-point amplitude~\cite{BRY,BDS, CompactThree}. Although this particular example is planar, it is just as simple to use the box cut for non-planar contributions.  For example, inserting the box into the four-point vertex in \fig{boxcutFigure} in a non-planar fashion generates non-planar four-loop integrals.

The box cut is an extremely efficient way to obtain contributions to parent graphs that contain a lower-loop four-point subgraph. As mentioned earlier, the  $L'=1$ box cut is closely related to the box-substitution rule depicted in~\fig{BoxSubRuleFigure}.  The box cut also generates contributions that are consistent with the two-particle cut and/or the rung rule~\cite{BRY}.

From the above covariant derivation, it follows that box cuts are valid in any dimension, if both the reduced cut $\tilde{{\cal A}}_n^{(L-L')}\big|_{\rm cut}$ and the four-point universal factor $\UniversalFactor^{(L')}$ are known in $D$ dimensions.  As a practical matter, the universal factors entering the lower-loop amplitudes should already be known in $D$ dimensions, prior to attempting the higher-loop calculation in $D$ dimensions. 

\section{Duality between color and kinematic factors}
\label{BCJ}

In modern S-matrix schemes, one successfully constructs multiparticle and multiloop amplitudes using minimal input from the theory as defined by the Lagrangian.  Rather than employing the nuts and bolts of the action, one relies heavily on quantum field theoretic consistency.  For example, a wealth of tree-level amplitudes for interesting theories can be fully constructed through BCFW~\cite{BCFW} using only the on-shell three-point amplitude as input.  Similarly, as discussed in \sect{MaximalCutsSection} and in greater detail in \cite{ZviYutinReview}, the unitarity method allows the construction of all loop level contributions  using as input only lower-loop subamplitudes.  A natural question is whether one can further minimize the theory-specific information needed to construct a full color-dressed scattering amplitude.  The conjectured duality between color and kinematic factors~\cite{BCJ, BCJloops} described in this section suggests such an approach. 

The duality relies upon the cubic-graph representation of gauge theory amplitudes, introduced earlier \eref{cubicRep}, which we write again as
\be
{\cal A}_n^{(L)}= i^L g^{n-2 +2L}\sum_{i\in\Gamma_3} \, \int \frac{d^{LD}\ell}{(2\pi)^{LD}} \frac{1}{S_i} \, \frac{N_i C_i}{p_{i_1}^2p_{i_2}^2p_{i_3}^2\cdots p_{i_{m}}^2}\,.
\label{LoopBCJ}
\ee
This time we have made the dependence on the coupling constant explicit, and pulled out an overall phase relative to \eref{cubicRep}.  We note that in general the $N_i$ cannot be uniquely specified;  if calculated by Feynman diagrams, they are gauge choice dependent.  If calculated by other means, they are allowed to transform under a generalized gauge transformation~\cite{BCJ, BCJloops,LagrangianSquare}, which corresponds to the ability to move contact terms between different graphs, while leaving the amplitude invariant.  This freedom allows for many different representations of the same amplitude, some of which make special properties of the theory manifest.

One very special property is the duality between color and kinematics. This duality asserts the existence of representations  \eref{LoopBCJ} where the kinematic numerators $N_i$ obey the same general algebraic properties as the color factors $C_i$. Specifically, the $N_i$ obey Jacobi relations and have antisymmetric cubic vertices analogous to the color factors, schematically:
\bea
 N_i+N_j+N_k=0~~ &\Leftrightarrow~~ C_i+C_j+C_k=0&\,,~~~~~~~~{\rm (Jacobi~identity)} \label{DefiningJacobi}  \\
~~~~~~~~~~N_{i}\rightarrow -N_{i}~ &\Leftrightarrow~~C_{i} \rightarrow -C_{i}\,& ~~~\mbox{(vertex-flip~antisymmetry)} \nn
\eea
where the first line signifies the Jacobi identity valid for specific triplets of graphs in the amplitude, and the second line represents the action of flipping the ordering of a cubic vertex in a graph.  In addition, it is often convenient at loop-level to impose the self-symmetries or graph automorphisms of the $N_i$, similar to the self-symmetries obeyed by the color factors $C_i$.

The duality between color and kinematics was first proposed by Bern and the present authors at tree-level~\cite{BCJ}, and later conjectured to hold to all loop orders~\cite{BCJloops} in generic gauge theories.  It generalizes a four-point property of tree-amplitudes observed in cuts of multi-loop-amplitudes~\cite{BCJ}, and also noticed in studies of zeros of scattering amplitudes in the 1980s~\cite{ZhuGHL}.  

Many of the the interesting and surprising consequences of this duality will not be treated in detail, as they are outside the scope of this review.   Nevertheless, in~\sect{TLGSection} we will briefly comment on some of these consequences, as related to tree amplitudes, Lagrangian manifestation, and constructing gravity amplitudes.  For the main body of this discussion, we will instead focus on how the duality can be used in the explicit construction of multiloop gauge theory amplitudes.  We will see that the constraints imposed by the duality allow us to drastically reduce the necessary input from unitarity cuts.  (As long as the duality remains a conjecture, it is of course necessary to verify any obtained amplitude on a set of spanning cuts.)  We may loosely quantify the amount of theory-specific information required for the construction of an amplitude by counting the number of necessary maximal and near-maximal cuts.  Alternatively, one may count the graphs used to represent the full amplitude and compare this number to the  very small number of graph numerators not completely fixed by the kinematic Jacobi relations~\eref{DefiningJacobi}. We call the latter ones ``master graphs''; from these, all other contributing graph numerators can be expressed using linear relations.  In \sect{ThreeLoopExampleSection}, the four-point  three-loop \NeqFoursYM\  amplitude, comprised of 12 graphs, is constructed using a single master graph.  The numerator of this single master graph, after imposing all symmetries and duality relations, is completely fixed by a single maximal cut, illustrating the dramatic simplification involved.

Nontrivial evidence supporting the duality conjecture at loop level is found by studying the four-point amplitude of the \NeqFoursYM\ theory at one~\cite{GSB}, two~\cite{BRY}, three~\cite{BCJloops}, and four loops~\cite{fourLoopBCJ}, and for the five-point amplitude through three loops~\cite{FivePointBCJ}.  Evidence that the duality holds even in nonsupersymmetric theories is provided~\cite{BCJloops} by the two-loop four-gluon identical-helicity pure Yang-Mills amplitude~\cite{BDKtwofourqcd}. 

We will elaborate on the duality using graph operators, but first it will be instructive to look at an venerable example from the literature where the duality \eref{DefiningJacobi} is known to hold.

%%%%%%%%%%%%%%%%%%%%
%FIGURE
\begin{figure}
\centerline{\epsfxsize 5 truein \epsfbox{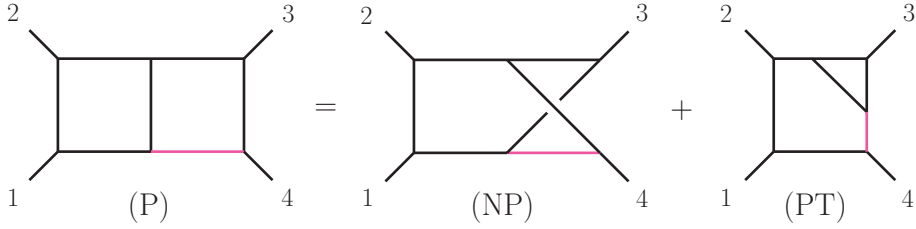}} \caption{ A Jacobi identity between three cubic four-point two-loop topologies.  These three graphs have the same connectivity except for the shaded (pink) edge.  Every Jacobi identity and kinematic identity between numerator factors can be graphically seen as a relation local to a set of three edges. For the \NeqFoursYM~amplitude, only graphs ${(\rm P)}$ and ${(\rm NP)}$ contribute~\cite{BRY}.  } \label{twoGraphs}
\end{figure}
%%%%%%%%%%%%%%%%%%%%%%

\subsection{Two-loop example} 

An instructive nontrivial example can be found using the four-point two-loop ${\cal N}=4$ sYM amplitude considered previously in \sect{MaximalCutsSection}.  As we have already shown how to construct the planar part using other methods, here we merely verify that the full amplitude, given in the literature{~\cite{BRY}}, satisfies the color--kinematics duality.  In the subsequent discussion we will show how such amplitudes can be directly constructed using the duality.
The two-loop amplitude is
\be
{\cal A}^{(2)}_{4}=-g^6 \Bigl(N^{(\rm P)}C_{1234}^{(\rm P)}I^{(\rm P)}(s,t) +N^{(\rm NP)}C_{1234}^{(\rm NP)} I^{(\rm NP)}(s,t)~+{\rm perms}\Bigr)\,,
\label{TwoLoopAmpEqn}
\ee
where the two scalar integrals contributing are drawn in~\fig{twoGraphs}, and the
``perms'' instructs us to sum over all distinct permutations of external labels. For convenience we have pulled the kinematic (and color) factors outside the integral, as they do not depend on the loop momenta. The numerators $N^{(\rm P)}$ and $N^{(\rm NP)}$ are given by
\be
N^{(\rm P)}=N^{(\rm NP)}= {\cal K} \, (k_1+k_2)^2 \,,
\label{Ndef}
 \ee
where ${\cal K}$ is the state-dependent prefactor introduced in \eref{Kprefactor}.

The color factors $C^{(\rm P)}$ and $C^{(\rm NP)}$, obtained by \eref{ColorDef},
are related through the Jacobi relation shown in \fig{twoGraphs}, to a penta-triangle graph $C^{(\rm PT)}$,
\be
C^{(\rm P)}=C^{(\rm NP)}+ C^{(\rm PT)}\,.
\ee
If \eref{TwoLoopAmpEqn} satisfies the color--kinematics~duality, we should expect a similar relation for the kinematic factors:
\be
N^{(\rm P)}=N^{(\rm NP)}+ N^{(\rm PT)}\,.
\label{NJacobi}
\ee
This indeed holds.  The amplitude \eref{TwoLoopAmpEqn} does not involve any penta-triangle integral; therefore, we set $N^{(\rm PT)}=0$, and thus \eref{NJacobi} is consistent with the result in \eref{Ndef}.

\subsection{Graph operators and the color--kinematics~duality}

In order to phrase the duality precisely, we will now introduce graph operators.  This is not an exercise in pedantry.  Indeed, precisely defining the duality in terms of graph operators allows for natural automation---especially useful when considering higher-loop contributions that may involve many graph topologies.  

\begin{figure}
\centerline{\epsfxsize 4.8 truein \epsfbox{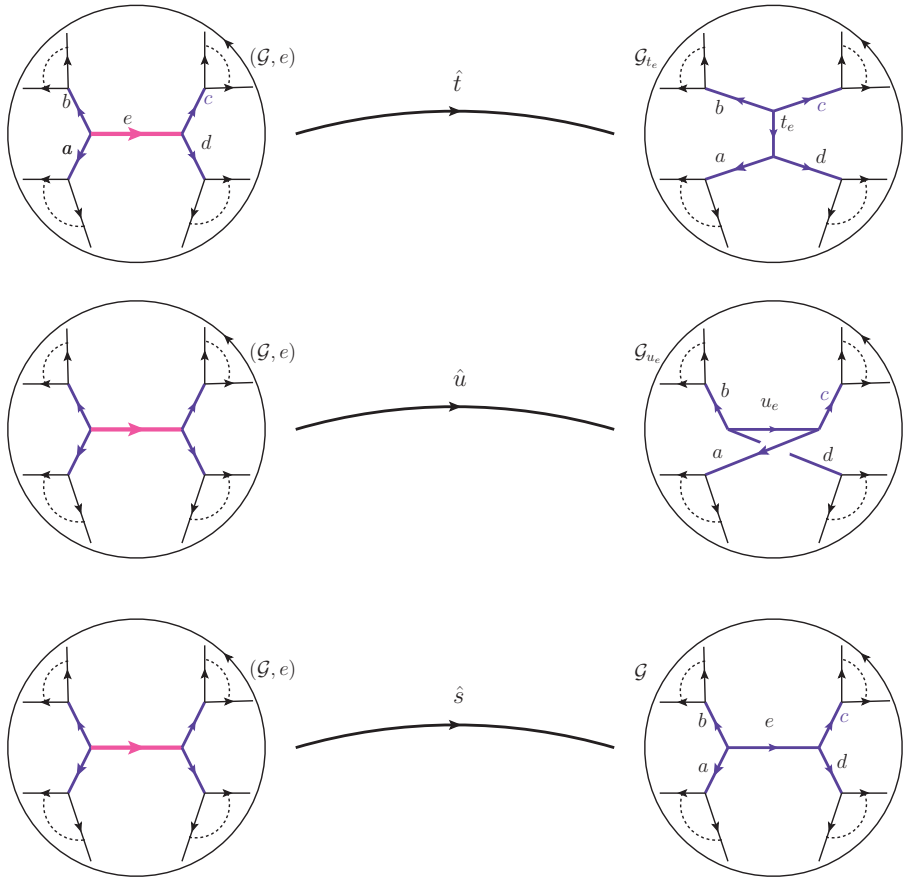}} \caption{Three graph operators taking $($graph ${\cal G}$ ,  edge $e$$)$ to graphs ${\cal G}_{t_e}, {\cal G}_{u_e}$ and ${\cal G}$.  All three graphs have the same connectivity, except for the edge touching edges ($a,b,c,d$).} \label{GraphOperators}
\end{figure}

The color Jacobi identity directly relates graphs that share entirely the same connectivity except for one edge, {\it e.g.} the shaded (pink) edge in \fig{twoGraphs}.    The language we will use to specify this relation will  involve operators that take a graph and a particular edge of that graph, and rearrange that edge's connectivity so as to get another graph.  We can specify any internal edge $e$  of a cubic graph in terms of the four edges that touch its bounding nodes.   If the order of edges around one node is $(a,b,e)$, and the order of edges around the other node is $(-e,c,d)$, we can specify the edge  $e$ as $e=\{(a,b),(c,d)\}$.   With this terminology we are ready to consider some mapping $\hat{t}$ to take a graph~${\cal G}$ and one of its edges $e$, to another graph ${\cal G}_{t(e)}$: 
\be
\hat{t}\big ({\cal G},\, e
\big ) = {\cal G}_{t(e)}\,,
\ee
where ${\cal G}_{t(e)}$ has the same connectivity of ${\cal G}$ except the nodes $(a,b,e)$ and $(-e,c,d)$ have been replaced with $(d,a,t)$ and $(-t,b,c)$, thereby defining the edge $t=\{(d,a),(b,c)\}$.   We will also use
\be
\hat{u}\big ({\cal G},\, e) = {\cal G}_{u(e)}\,,
~~~~~~~
\hat{s} \big ({\cal G},\, e \big )= {\cal G}\,,
\ee
where ${\cal G}_{u(e)}$ has the same connectivity of ${\cal G}$ except nodes $(a,b,e)$ and $(-e,c,d)$ have been replaced with $(d,b,u)$ and $(-u,c,a)$, with edge $u=\{(c,a),(d,b)\}$.  The $\hat{s}$ operator simply returns the original graph unchanged independent of the edge specified.   These operations are depicted visually in \fig{GraphOperators}.  

%%%%%%%%%%%%%%%%%%%%
%FIGURE
%%%%%%%%%%%%%%%%%%%%%%

We are now ready to concisely specify the color Jacobi identity associated with every edge $e$ of every graph $\cal G$,
\bea
0&=  C \left[ \hat{s} \big ( {\cal G}, e \big )\right]-C \left[ \hat{t} \big ( {\cal G}, e \big )\right] -C \left[ \hat{u} \big ( {\cal G}, e \big )\right] \nn \\
&=C[{\cal G}]- C[ {\cal G}_{t(e)}] -C[ {\cal G}_{u(e)}]  \,,
\label{colorJacobi}
\eea
where $C[{\cal G}]$ gives the color factor arrived at by dressing every node of graph ${\cal G}$ with the $\f^{abc}$ structure constants. The color--kinematics duality is satisfied in a representation where for all edges $e$ of all graphs $\cal G$,
\be
\label{numIdentity}
0= N \left[ \hat{s} \big ( {\cal G}, e \big )\right]-N \left[ \hat{t} \big ( {\cal G}, e \big )\right] -N \left[ \hat{u} \big ( {\cal G}, e \big )\right] \,,
\ee
 where $N[{\cal G}]$ gives the kinematic numerator factor associated with the graph ${\cal G}$. 

Additionally, we may use this formalism to write down some relations that originate from the antisymmetry of the cubic vertices
\bea
C[{\cal G}_{\{(d,a),(b,c)\}}]&=&-C[{\cal G}_{\{(a,d),(b,c)\}}]\,,\nn \\
N[{\cal G}_{\{(d,a),(b,c)\}}]&=&-N[{\cal G}_{\{(a,d),(b,c)\}}]\,.
\label{antisym}
\eea
In the following section, we will see that the set of constraints on the numerator factors \eref{numIdentity} and \eref{antisym} can be understood as a set of functional equations that must hold for the duality to be satisfied.

\subsection{Three-loop example}
\label{ThreeLoopExampleSection}
%%%%%%%%%%%%%%%%%%%%
%FIGURE
\begin{figure}[t]
\centerline{\epsfxsize 4.5 truein \epsfbox{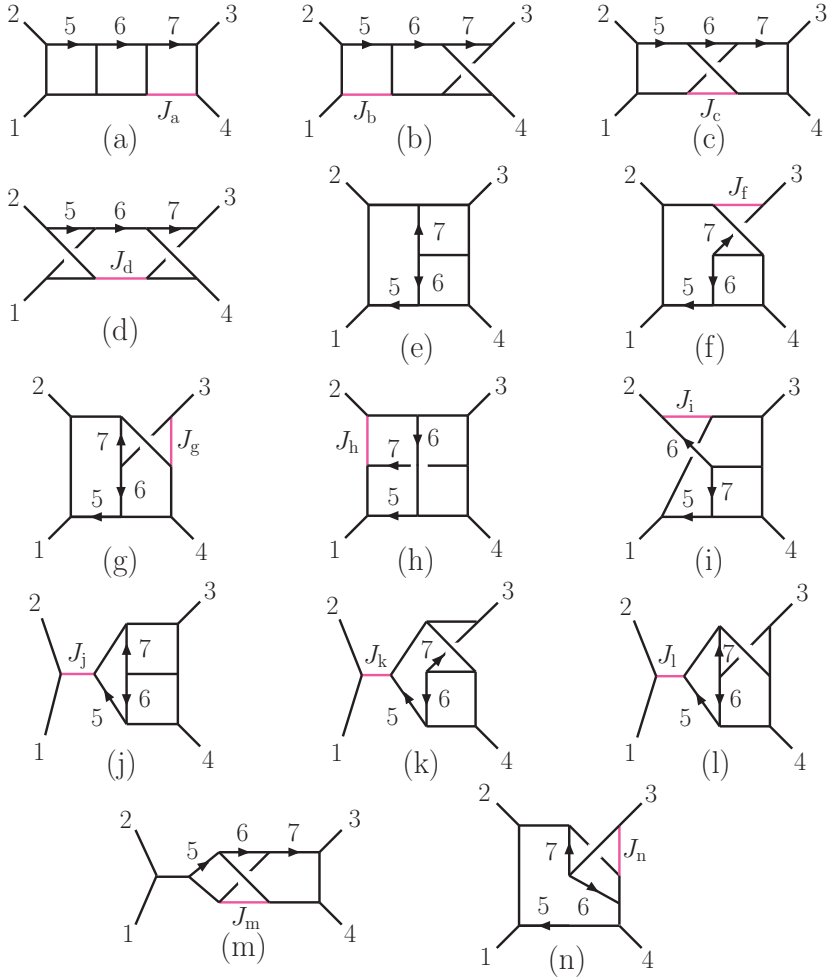}} \caption{Three-loop four-point cubic graphs considered in the main text. The external momenta is outgoing and the shaded (pink) edges mark the application of kinematic Jacobi relations used in  \eref{BCJjacobi}. Note that only graphs (a)--(l) contribute to the \NeqFoursYM~amplitude where the duality between color and kinematics is made manifest.}
\label{threeLoop14}
\end{figure}
%%%%%%%%%%%%%%%%%%%%%%

In this section, we reexamine the four-point three-loop \NeqFoursYM\ amplitude using the duality between color and kinematics~\cite{BCJloops}.  This amplitude was  originally given in \cite{superfinite,CompactThree} in terms of nine cubic diagrams.  For this exercise, we start by considering a larger set of 25 graphs, which are related to any of the original nine diagrams by a single application of a kinematic Jacobi relation. However, eleven of these diagrams contain triangle subgraphs, which the no-triangle property of \NeqFoursYM~\cite{MHVoneloop} suggests will not contribute.  After removing those with one-loop triangle subgraphs we have the 14 graphs depicted in \fig{threeLoop14}. We will see that this set of diagrams is sufficiently large to admit a manifest representation of the duality. 

Now we will introduce the kinematic Jacobi relations that the numerators of each diagram must satisfy.  Each numerator depends on three independent external momenta and (at most) three independent loop momenta.  While in general a graph's numerator factor will depend on the states of all external particles, the four-point amplitude of \NeqFoursYM\ is special in that all of its state dependence can be captured in the overall factor ${\cal K}$ described in \sect{RecursiveCutsSection}. Since this factor is common to all terms and the kinematic Jacobi relations are homogeneous, this factor plays no role in the equations.  With this in mind, we suppress all state dependence but make the dependence on momenta explicit:
\be
N^{(x)}=N^{(x)}(k_1,k_2,k_3,l_5,l_6,l_7)\,,
\ee
where the canonical labeling for each  graph $x\in \{\rm a,b,c,d,e,f,g,h,i,j,k,l,m,n\}$ is given in \fig{threeLoop14}.

For every internal edge and for every graph there is one functional equation relating the numerator of that graph to the numerator of two other graphs through the kinematic Jacobi relations.   The $14$ graphs of \fig{threeLoop14} each have ten propagators, giving 140 relations in total (before accounting for possible overcounts). One can write down many more equations using the graph automorphisms of each numerator -- ensuring that numerators are invariant under equivalent ways of labeling symmetric graphs.  However, this large set of equations contains much redundancy; the number of independent functional relations is quite small. We will not elaborate on how to methodically reduce this system; instead, we observe that the following 13 kinematic Jacobi relations, together with all automorphism relations, solve all constraints imposed by the color--kinematics duality:

 \be
 \begin{array}{lc}
 N^{(\rm a)}= N^{(\rm b)}(k_1,k_2,k_3,l_5,l_6,l_7)\,, & (J_{\rm a })\\
 N^{(\rm b)}= N^{(\rm d)}(k_1,k_2,k_3,l_5,l_6,l_7)\,, &  (J_{\rm b })\\
 N^{(\rm c)}= N^{(\rm a)}(k_1,k_2,k_3,l_5,l_6,l_7)\,, & (J_{\rm c })\\
 N^{(\rm d)}= N^{(\rm h)}(k_3,k_1,k_2, l_7, l_6, k_{1,3}-l_5+l_6-l_7)&\\
~~~~~~~~~\null+N^{(\rm h)}(k_3, k_2, k_1,l_7,l_6,  k_{2,3}+l_5-l_7)\,,& (J_{\rm d })\\
 N^{(\rm f)}= N^{(\rm e)}(k_1,k_2,k_3,l_5,l_6,l_7)\,, & (J_{\rm f })\\
 N^{(\rm g)}= N^{(\rm e)}(k_1,k_2,k_3,l_5,l_6,l_7)\,, & (J_{\rm g })\\
 N^{(\rm h)}= - N^{(\rm g)}(k_1, k_2, k_3, l_5, l_6, k_{1,2} - l_5 - l_7)&\\
~~~~~~~~~\null-N^{(\rm i)}(k_4, k_3, k_2,  l_6-l_5,  l_5 - l_6 + l_7-k_{1 ,2}, l_6)\,, & (J_{\rm h })\\
 N^{(\rm i)}\,= N^{(\rm e)}(k_1,k_2,k_3,l_5,l_7,l_6)&\\
~~~~~~~~~\null-N^{(\rm e)}(k_3, k_2, k_1, -k_4 - l_5 - l_6, - l_6 - l_7, l_6)\,, & (J_{\rm i })\\
 N^{(\rm j)}\,= N^{(\rm e)}(k_1,k_2,k_3,l_5,l_6,l_7)-N^{(\rm e)}(k_2,k_1,k_3,l_5,l_6,l_7)\,,& (J_{\rm j })\\
 N^{(\rm k)}= N^{(\rm f)}(k_1,k_2,k_3,l_5,l_6,l_7)-N^{(\rm f)}(k_2,k_1,k_3,l_5,l_6,l_7)\,, & (J_{\rm k })\\
 N^{(\rm l)}\,= N^{(\rm g)}(k_1,k_2,k_3,l_5,l_6,l_7)-N^{(\rm g)}(k_2,k_1,k_3,l_5,l_6,l_7)\,, & (J_{\rm l })\\
 N^{(\rm m)}= 0\,, & (J_{\rm m })\\
 N^{(\rm n)}\,=0\,, & (J_{\rm n})
 \label{BCJjacobi}
 \end{array}
 \ee
where $k_4=-(k_1+k_2+k_3)$,  $k_{i,j}=k_i+k_j$, and  for notational simplicity we have suppressed the arguments of the numerators on the left-hand-sides;  these all have the  canonical argument $(k_1,k_2,k_3,l_5,l_6,l_7)$.  The 13 relations are labeled by $(J_x)$ which correspond to Jacobi relations specified by the marked edges in \fig{threeLoop14}. In the above relations we have immediately set to zero any diagrams that do not feature in \fig{threeLoop14}. This explains why some relations in \eref{BCJjacobi} involve fewer than three terms.
 
 Analyzing the result of \eref{BCJjacobi}, we immediately see that the color--kinematics~duality has constrained $N^{(\rm m)}$ and $N^{(\rm n)}$ to vanish, given that we insist on the absence of graphs with one-loop triangles.  
 We see that $N^{(\rm f)},N^{(\rm g)},N^{(\rm i)},N^{(\rm j)}$ are directly expressed as linear combinations of the function $N^{(\rm e)}$. Furthermore, $N^{(\rm k)},N^{(\rm l)},N^{(\rm h)}$ are expressed as linear combinations of the functions  $N^{(\rm f)},N^{(\rm g)},N^{(\rm i)},N^{(\rm j)}$, and thus they can also be solved in terms of $N^{(\rm e)}$. Finally we note that $N^{(\rm a)}=N^{(\rm b)}=N^{(\rm c)}=N^{(\rm d)}$ and that $N^{(\rm d)}$  can be expressed as a linear combination of $N^{(\rm h)}$.  So each of these numerators can also be solved in terms of $N^{(\rm e)}$. Therefore, every numerator $N^{(x)}$ is entirely expressible in terms of the planar diagram numerator $N^{(\rm e)}$.
 
Although the solution to \eref{BCJjacobi} is quite constraining, it does not enforce all of the 140 Jacobi-like relations we began with, nor does it enforce all automorphism relations. As mentioned, we will obtain the full solution by enforcing automorphic invariance of each graph numerator.  Using the partial solution of \eref{BCJjacobi}, these relations can be translated to self-constraints on $N^{(\rm e)}$.   For example, the \fig{threeLoop14}(e) diagram is invariant under reflections up $\leftrightarrow$ down; thus, $N^{(\rm e)}$ must satisfy
\be
N^{(\rm e)}(k_1,k_2,k_3,l_5,l_6,l_7) =N^{(\rm e)}(k_2,k_1,k_4,k_{1,2}-l_5,l_7,l_6)\,.
\ee 
Similarly, \fig{threeLoop14}(j) is invariant under $1\leftrightarrow2$. However, just like color factors, numerators are odd under the reordering of a cubic vertex; therefore, we have
\be
N^{(\rm j)}(k_1,k_2,k_3,l_5,l_6,l_7) =-N^{(\rm j)}(k_2,k_1,k_3,l_5,l_6,l_7) \,.
\ee 
In fact, the reader may check that this equation is already enforced by \eref{BCJjacobi}; thus, this particular automorphism relation does not give rise to any new constraints on $N^{(\rm e)}$.
Beyond these two examples, every diagram but  \fig{threeLoop14}(g) is constrained by automorphism relations, which through \eref{BCJjacobi} translates to possible further constraints on $N^{(\rm e)}$. 

One way to efficiently solve the remaining kinematic Jacobi and automorphism relations is to write down an ansatz for $N^{(\rm e)}$. As mentioned, we assume that the state-dependent prefactor is ${\cal K}\equiv s_{12} s_{23} A_4^{\rm tree}(1,2,3,4)$.   We write the remaining factors as a local degree-two polynomial in 17 momentum invariants:
\be
N^{(\rm e)}_{\rm ansatz}={\cal K} \, P(s_{12},s_{23},k_i\cdot l_j,l_i\cdot l_j)\,,
\ee
where $k_i=k_1,k_2,k_3$ and  $l_i=l_5,l_6,l_7$, and the degree of the polynomial is given by simple dimensional counting. Although we could use various facts about the power counting of \NeqFoursYM\  to simplify the ansatz, we proceed with the most general homogeneous degree-two polynomial of 17 variables, which gives $17 \times(17+1)/2=153$ parameters to be determined. 

Using such an ansatz, we find that only the following four automorphism relations are required in order for all automorphic relations to be satisfied:
\bea
N^{(\rm c)}(k_1,k_2,k_3,l_5,l_6,l_7)&=&N^{(\rm c)}(k_1,k_2,k_4,l_5,l_5-l_6,k_{3,4}-l_7)\nn \,, \\
N^{(\rm f)}(k_1,k_2,k_3,l_5,l_6,l_7)&=&N^{(\rm f)}(k_2,k_1,k_4,k_{1,2}-l_5,k_4-l_5-l_7,k_4+l_5-l_6)\nn \,, \\
N^{(\rm h)}(k_1,k_2,k_3,l_5,l_6,l_7)&=&N^{(\rm h)}(k_1,k_4,k_3,k_1 -l_5,l_7,l_6)\nn\,, \\
N^{(\rm h)}(k_1,k_2,k_3,l_5,l_6,l_7)&=&N^{(\rm h)}(k_2,k_1,k_4,k_{1,2}-l_5-l_7,-l_6,l_7) \,.
\label{automorph}
\eea
Enforcing  \eref{BCJjacobi} and \eref{automorph},  the ansatz is constrained from 153 parameters to only three free parameters, which we label $\alpha, \beta, \gamma$, giving
\bea
N^{(\rm e)}_{\rm sol}=&{\cal K} \, \Big[ \alpha\, s_{12}s_{23}  +
\beta \, ( s_{12}^2  +  s_{23}^2)  + \gamma \, \Big(s_{12}^2+
2 (s_{12}- s_{23})\, k_1 \cdot l_5  \nn \\& \hskip 3.5cm
+2 (2 s_{12} +s_{23})\, k_2 \cdot l_5  + 6 s_{12} \, k_3 \cdot l_5 \Big) \Big]\,.
\label{BCJsolution}
\eea
After checking that this also solves the original 140 Jacobi relations and all automorphisms, we conclude that \eref{BCJsolution} is a threefold family of potential solutions that makes the duality between color and kinematics manifest. 

We still need to ensure that these solutions actually describe the ${\cal N}=4$ theory.  One such source of theory-specific data can be a generalized unitarity cut.  We consider the maximal cut of graph (e), using the known expression in the literature~\cite{BRY}, 
\be
N^{(\rm e)}\Bigl|_{\rm cut}=2 \, {\cal K} \, s_{12}\,k_4 \cdot l_5 \,.
\label{MCut3L}
\ee
Then, we compare to the maximal cut of the solution \eref{BCJsolution},
\be
N^{(\rm e)}_{\rm sol}\Bigl|_{\rm cut}={\cal K} \, \Big[ \alpha\, s_{12}s_{23}  +
\beta \, ( s_{12}^2  +  s_{23}^2)  + \gamma \, (
s_{12}s_{23}  - 6 s_{12} \, k_4 \cdot l_5 ) \Big]\,,
\ee
where we have used the relations $k_1\cdot l_5=0$, $ 2 k_2\cdot l_5=s_{12}$, and $ 2 k_3\cdot l_5=-s_{12}-2 k_4\cdot l_5$, valid on the maximal cut.

Comparing \eref{BCJsolution} and \eref{MCut3L}, we find that  $\alpha=1/3$,
$\beta=0$, $\gamma=-1/3$, giving the fully solved numerator
\bea
N^{(\rm e)}&=&\frac{1}{3}{\cal K} \, \Bigl( s_{12}(s_{23}-s_{12})
-2 (s_{12}- s_{23})\, k_1 \cdot l_5 \nn \\&&
~~~~~~~  -2 (2 s_{12} +s_{23})\, k_2 \cdot l_5  - 6 s_{12} \, k_3 \cdot l_5  \Bigr)\,.
\label{FullBCJsolution}
\eea
All 12 nonvanishing numerators can now be obtained from \eref{FullBCJsolution}, using  \eref{BCJjacobi}. They are explicitly given in~\cite{BCJloops}.  Together, they represent a duality-satisfying expression of the three-point four-loop \NeqFoursYM~amplitude.  This was verified on a spanning set of cuts in~\cite{BCJloops}. 
 
\subsection{Tree level, Lagrangians, and gravity}
\label{TLGSection}

While this review focuses on multiloop methods for gauge theory amplitudes, we should mention a few other interesting results intimately related to the kinematics--color duality.
Consider the formula \eref{LoopBCJ} in the tree-level limit $L=0$.  This gives the cubic representation of tree amplitudes introduced in~\cite{BCJ}:
\be
{\cal A}_n^\tree= g^{n-2}\sum_{i\in\gamma_3} \, \frac{n_i c_i}{p_{i_1}^2p_{i_2}^2p_{i_3}^2\cdots p_{i_{m}}^2}\,,
\ee
where the sum runs over all cubic tree-level graphs $\gamma_3$ (including all distinct permutations of external legs), which all have unit symmetry factors.  For notational uniformity with~\cite{BCJ} we use $n_i=N_i^\tree$ and $c_i=C_i^\tree$.  

As shown in~\cite{BCJ}, assuming the duality between tree-level color and kinematics leads to novel relations between color ordered tree amplitudes. These decompose any color-ordered partial amplitude into a sum of $(n-3)!$ basis amplitudes, schematically,
\be
A_n^{\tree}( \sigma(1),\sigma(2),\ldots, \sigma(n))= \sum_{\rho  \in S_{n-3}} K^{(\sigma)}_{\rho} A_n^{\tree}(1,2,3,\rho(4),\ldots, \rho(n))\,,
\label{amplituderelations}
\ee
where $\sigma$ and $\rho$ are permutations, and $K^{(\sigma)}_{\rho}$ are some universal (state-independent) non-local kinematic coefficients. We will not give these factors here; they can be deduced using the results in~\cite{BCJ}. 
Similar relations between partial amplitudes were subsequently found in string theory~\cite{Monodromy} using monodromy of integration on the worldsheet. In the field-theory limit, these string relations provided a first proof of \eref{amplituderelations}. Using BCFW recursion, the same relations were later derived from field theory~\cite{amplituderelationProof}. In \cite{noncommutative}, analogous relations was shown to exist in non-commutative Yang-Mills theory. 

As for the duality itself, phrased in terms of color and kinematic numerator factors and corresponding Jacobi relations~\eref{DefiningJacobi}, less is known.  An investigation in this direction is \cite{TyeDoubleCopy} where the framework of heterotic string theory is used to shed light on the duality.   In addition, there are known explicit solutions for tree-level duality-satisfying numerators $n_i$ to all multiplicity~\cite{KiermaierTalk, BBDSVsolution}. While these solutions demonstrate the existence of numerators satisfying the duality, they take a particularly non-local and asymmetric form (under relabeling of external states). The existence of local or symmetric forms of duality-satisfying tree-level numerators to all mulitplicity remains an open question.

In~\cite{LagrangianSquare}, it was shown that the duality between color and kinematics could be made manifest at the Lagrangian level of Yang-Mills theory, by introducing higher degree interactions at the five-point and six-point levels, schematically:
\be
{\cal L}_{\rm YM}=\frac{1}{4g^2}F^{\mu \nu}F_{\mu \nu}+ {\cal L}_5+ {\cal L}_6 + \ldots \, ,
\ee
and at the same time introducing auxiliary fields so to make all interactions cubic. Remarkably the corrections ${\cal L}_5+ {\cal L}_6 + \ldots$ vanish by the usual gauge-group Jacobi identity, and do not alter the theory in any  way other than modifying the relative values of individual Feynman diagrams. 

Finally, we remark that one can easily construct gravity amplitudes using the duality; gravity numerators are double copies of gauge theory kinematic factors~\cite{BCJ,BCJloops}. Given an $n$-point  $L$-loop gauge theory amplitude with duality satisfying numerators $N_i$, a gravity amplitude takes the form
\be
{\cal M}_n^{(L)}= i^{L+1} \left(\frac{\kappa}{2}\right)^{n-2 +2L}\sum_{i\in\Gamma_3} \, \int \frac{d^{LD}\ell}{(2\pi)^{LD}} \frac{1}{S_i} \, \frac{N_i \widetilde N_i}{p_{i_1}^2p_{i_2}^2p_{i_3}^2\cdots p_{i_{m}}^2}\,,
\label{GravBCJ}
\ee
where  $\tilde{N}_i$ are a set of  kinematic numerators for the amplitude of a possibly different gauge theory (which need not explicitly satisfy the duality), and $\kappa$ is the gravitational coupling constant. This formula is the generalization of the analogous tree-level ($L=0$) formula~\cite{BCJ}. Assuming the color--kinematics duality, the double copy property was proven~\cite{LagrangianSquare}  recursively in $n$ for the explicit case of maximal supersymmetry
$[\NeqFour] \otimes [\NeqFour] \rightarrow [\NeqEight]$ and similarly for pure Yang-Mills theory going to pure gravity plus axion and dilaton.  At the classical level,  \eref{GravBCJ}  encodes (and arguably explains) the field-theory forms of the string-theoretic Kawai-Lewellen-Tye (KLT) tree-level relations~\cite{KLT} between gravity scattering amplitudes and gauge theory amplitudes.   Intriguingly, the double copy form of gravity \eref{GravBCJ} appears  to generalize seamlessly to the quantum level, to any loop order~\cite{BCJloops}. 

We close this section by referring the reader to the literature for further application of the duality, the new tree-level amplitude relations, and the gravity double copy property~\cite{BCJOther,Neq44np,BBDSVsolution,Mafra};  see also reviews~\cite{BCJreview}.

\section{Conclusions}

In this review we have covered some recent methods for constructing compact integral representations of multiloop scattering amplitudes, applicable to planar and non-planar gauge theory in general, and to \NeqFoursYM\ theory in particular, valid in four dimensions and in $D>4$. 

The maximal cut method is a convenient scheme in the generalized unitary framework that allows the construction of amplitudes in any massless theory starting from the cuts with the most on-shell propagators. The method has a natural hierarchy which determines the order in which unitary cuts should be evaluated in order to capture missing contributions of an initial amplitude ansatz. Maximal cuts are followed by next-to-maximal cuts and thereafter next-to-next-to-maximal cuts and so on, until the full amplitude has been determined. These maximal and near-maximal cuts can be evaluated in four dimensions and in $D>4$, for planar as well as non-planar cuts. A close relative to the maximal cut method is the leading singularity method~\cite{CachazoSkinner,CachazoLeadingSingularityAndCalcs,Spradlin3loop}, which focuses on the \NeqFoursYM\ theory in four dimensions. 

Complementary to the maximal cut method, we have reviewed two particularly simple $D$-dimensional multiloop unitarity cuts in the \NeqFoursYM\ theory: the ordinary two-particle cut of the four-point amplitude, and the box cut. These two types of cuts have a recursive structure that can be implemented using simple pictorial calculations.

Although the \NeqFoursYM\ theory naturally lives in four dimensions, the $D$-dimensional cuts and amplitudes are important for several reasons.
Since \NeqFoursYM\  amplitudes exhibit infrared divergences at loop level they need to be regulated, either using dimensional regularization, or using mass regulators~\cite{MassiveRegulatorProgress}. The former method naturally requires $D$-dimensional amplitudes, and the latter can in many cases be rephrased in terms of $D$-dimensional amplitudes, where the higher-dimensional momenta effectively act as masses from a four-dimensional perspective. Beyond this, there is, of course, the inherent theoretical interest in studying higher-dimensional theories.

Using the duality between color and kinematic factors we considered how to compute multiloop integrands in a very efficient manner. We discussed the three-loop four-point amplitude of \NeqFoursYM\ in some detail, showing how the duality can be made manifest using 12 nonvanishing cubic three-loop graphs, all determined from only one master graph. The duality, which is conjectured to hold for any multiparticle and multiloop amplitude in \NeqFoursYM, as well as less-supersymmetric theories (including pure Yang-Mills), hints of some deep field-theoretic origin, which once understood will explain the suggestive new ways to organize and calculate scattering amplitudes. Similarly, the connection to gravity amplitudes as double copies of gauge theories, similar to the string-theoretic KLT relations, hints of an underlying unifying framework that includes a particularly novel notion of graviton compositeness.  From this perspective a connection to string theory is not surprising.

\ack
We are grateful to Zvi Bern, Camille Boucher-Veronneau, Johannes Br\"odel, Tristan Dennen, Lance Dixon, Daniel Freedman, Yu-tin Huang, Harald Ita, Renata Kallosh, David Kosower, and Radu Roiban for collaboration on these very topics, valuable comments on this manuscript,  and/or many stimulating discussions on related subjects.  JJMC gratefully acknowledges the Stanford Institute for Theoretical Physics for financial support. HJ's research is supported by the European Research Council under Advanced Investigator Grant ERC-AdG-228301.
The figures were generated using Jaxodraw~\cite{Jaxo1and2}, based on Axodraw~\cite{Axo}.

\end{document}